\documentclass[12pt]{iopart}

\usepackage{graphicx}

\def\pr{\mathop{\prime}\nolimits}

\def\t{\mathop{\vartheta}\nolimits}

\def\p{\mathop{\varphi}\nolimits}
\def\r{\mathop{\varrho}\nolimits}
\def\o{\mathop{\omega}\nolimits}
\def\c{\mathop{\chi}\nolimits}
\def\l{\mathop{\lambda}\nolimits}
\def\a{\mathop{\alpha}\nolimits}
\def\b{\mathop{\beta}\nolimits}

\def\m{\mathop{\mu}\nolimits}

\def\Bb{\mathop{\mbox{\boldmath$\beta$}}\nolimits}
\def\Bn{\mathop{\mbox{\boldmath$\nabla$}}\nolimits}
\newcommand{\bm}[1]{\mbox{\boldmath$#1$}}

\begin{document}
\title[]{Selfconsistent Numerical Calculation of Relativistic Neutron Star Magnetospheres}


\author{Peter Biltzinger and K.O. Thielheim}
\address{University of Kiel, Department of Physics, 24118 Kiel, Germany}

\begin{abstract}
Selfconsistent magnetospheres of rotating cosmic magnets (neutron stars/pulsars) with arbitrary inclination of the magnetic against the rotation axis are considered. Present studies concentrate
on the regime dominated by the force--free surface (FFS). A macroscopic fluid description is applied and radiation reaction is taken into account. As in earlier work of our group, a 'standard set of parameters' is used. Under these conditions, the following features are found among other results: global charge separation exists for all degrees of inclination of the magnetic against the rotation axis; clouds of different charge are seperated by regions of vanishing particle number density; as expected, test particles inserted into the latter regions propagate into one of the adjacent clouds; strong polodial currents exist; locally averaged particle energies for protons typically range up to $10^{16}-10^{17}$ eV, depending on the angle of inclination.
\end{abstract}



\maketitle

\section{Introduction}
Since the discovery of pulsars in 1968 \cite{hewish} and their interpretation as a rapidly rotating magnetized neutron stars in the same year \cite{gold, pacini}, these compact objects are under discussion as powerful accelerators of ultra high energy cosmic ray particles.

From the very beginning of neutron star physics, work on the dynamics of electrically charged particles accelerated in the corresponding electromagnetic fields proceeded on two stages: (1) test particle dynamics in the vacuum fields of rotating magnets, e.g.~\cite{ostriker, gunn}, demonstrating fundamental mechanisms and (2) self consistent plasma dynamics, e.g.~\cite{goldreich}, reproducing certain aspects of the structure and evolution of neutron star magnetospheres.

Since then, numerous papers have been published, investigating these matters in great detail, which we will not be able to discuss or even just to mention in this introduction to our present paper. Here we shall concentrate on stage (2), on relativistic plasma dynamics in a regime governed by the force--free surface of a homogeneously magnetized, rapidly rotating sphere with parameters typical for neutron stars.

The notation of a force--free surface (FFS) refers to the dynamic of an electrically charged (test) particle within given electromagnetic fields. By definition the FFS is generated by those points of configuration space at which the Lorentz--force acting on that particle through given electromagnetic fields vanishes. While in published literature a particle that happens to be at such a point often referred to as 'force-free' (ff), we prefer -- in view of the presence of other types of electromagnetic forces (radiation reaction forces) -- to speak of a Lorentz--force--free (Lff) particle in that situation. If ${\bf B} \neq 0$ and ${\bf E} \neq 0$ \footnote{Boldface letters are for vectors in 3-dimensional euklidean space. $(\cdot,\cdot)$ denotes the scalar product, $[\cdot,\cdot]$  the vector product}, as is the case in fields considered here, a particle is  Lff for ${\bf E} + [\Bb , {\bf B}] = 0$, i.e., for (I) $(\bf{E}, \bf{B}) = 0$ and (II) $(\bf{E}, \Bb)= 0$ and (III) $([\bf{E},\bf{H}], \bm{\beta}) = |\bf{E}^2|$. ${\bf E}$ is the electric field vector, ${\bf B}$ the vector of the magnetic induction  and $\Bb$  is the velocity in units of the velocity of light. For $| \bf{E} | \ll | \bf{B} |$, as is the case under premises adopted here, one may expect the second and the third conditions to be inherently fulfilled to some approximation so that the FFS then is caracterized solely by the first condition, $(\bf{E}, \bf{B}) = 0$.

From the early works of \cite{jack1, jack2} magnets rotating in the vacuum with the vector of magnetic dipole moment inclined against their respective rotation axis are known to create such FFS, of which some segments can act as particle traps and thus may have strong bearings on the formation of a neutron star magnetosphere, at least within a certain range of distance from its surface.

In what follow, it will be useful to distinguish the special case of {\em aligned rotators}, i.e.~rotating magnets with the magnetic axis parallel to the rotation axis ({\em parallel rotators}) or antiparallel to the rotation axis ({\em antiparallel rotators}), from the general case of {\em inclined rotators} and from rotating magnets with the magnetic axis orthogonal to the rotation axis ({\em orthogonal rotators}). On stage (2), a considerable number of investigations has been published on aligned rotators. Some of these papers will be mentioned below. But only few are devoted to inclined or orthogonal rotators. In the latter case, obviously, considerable formal and numerical complications arise from the lack of rotational symmetry. Analytical approaches were used, for example, by \cite{jack2, endean, uchida, shibata}.

In the special case of aligned rotators, due to axial symmetry, theoretical results are achieved much easier, even on stage (2). Analytical methods have been applied to the structure of the magnetosphere of aligned rotators for example by \cite{goldreich, jack1, mestel, fritz1, fritz2, kabu1, kabu2, rylov1, rylov2}. Also, numerical studies on that matter have been performed by \cite{kuo1, kuo2, krause, ertl, zachariades, neukirch}.

On stage (1) in an earlier work of our group \cite{wolf89} the authors have integrated numerically the equation of motion for individual test particles within the regime of the FFS and with no restrictions on the relative orientation of the rotational towards the magnetic axis. Results confirm that velocity components orthogonal to the magnetic field vector are efficiently damped by radiation reaction. Thus, test particles tend to follow magnetic field lines, as suggested earlier by \cite{jack1}. In a certain class of orbits they oscillate about the FFS, while moving along magnetic field lines. The amplitude of these oscillations decreases through radiation losses. Ultimately, in the subsequent regime of lower energy (and on a much larger time scale) particles become subject to drift in azimuthal direction.

On stage (2) in a second paper of our group \cite{wolf94} the authors have introduced a numerical iterative approach to reproduce sequences of quasi stable plasma configurations forming under the influence of the FFS. In iterative steps charged particles were allowed to be ejected from surface elements of the rotating sphere in quantities locally proportional to the magnitude of the electric vector component normal to the respective surface element. These particles were then allowed to move freely along the appropriate magnetic field lines and to settle down where the projection of the electric onto the magnetic vector vanishes. Thus, {\em without making use of the equation of motion}, co-rotating, quasi stable, charge separated clouds were reproduced, in consistency with earlier results mentioned above.

In our present work we proceed one step further taking into account in a selfconsistent way virtually {\em all effects of relativistic particle dynamics}, including radiation reaction and effects of special relativity as, for example, retardation. Our numerial approach is designed  to describe the evolution of locally averaged particle densities, since velocity dispersion is not taken into account. Again, a magnetosphere is allowed to build up from the initial vacuum through particle ejection from the spherical surface, similar to the procedure described above. Here particles are allowed to be ejected from surface elements, given an appropriate direction of the electric vector,in quantities locally proportional to the magnitude of the electric vector component projected onto the magnetic field line.

Thereby, we intend to clearify, on stage (2), the evolution, selfconsistent structure and stability properties of plasma configurations forming within the regime of the FFS of a rotating cosmic magnet with no restrictions on the relative orientation of the rotational towards the magnetic axis. Also, we want to evaluate mean energy values locally achieved by particles in that regime.

As in earlier works of our group, e.g.~\cite{laueth, th91, th94}, we  apply a model represented by a rotating, ideally conducting, homogeneously magnetized sphere with arbitrary inclination of the magnetic against the rotation axis. A 'standard set of parameters' representing well-known properties of typical neutron stars is attributed to this model: the stellar mass which is taken equal to the solar mass $m_{N} =  m_{{\rm sun}}$, the stellar radius $r_{N} = 10^6$cm, the angular velocity $\omega = 20\,\pi\,{\rm s}^{-1}$, and the magnetic dipole moment $\mu = 10^{30}$ G ${\rm cm}^3$ \footnote{Gaussian units are used throughout this paper. Thus, electric and magnetic field strengths are measured in units of $1 {\rm G} = 300 {\rm V/cm}$. For the standard set of parameters, the magnetic field strength is about $B_{p} \simeq 2 \cdot 10^{12}$ \,G and the electricic field strength is approximately $E_p \simeq 10^{10} G$ in the polar region. Under given parameter values, the radius of the light--cylinder (often referred to as 'light radius') $r_L = \o/c$, outside of which corotation cannot exist, is $r_L = 5 000$ km, corresponding to almost the radius of the earth, for comparison.}.

In order to investigate the regime of the FFS with an appropriate resolution we concentrate on the near zone of the neutron star up to $20\,r_N$.

From preceeding estimates as well as from subsequent simulations gravitational forces exerted by the rotating neutron star and by the magnetosphere itself onto individual particles, as well as effects of general relativity were found to be negligible for the standard set of parameters. Contributions to the  Lorentz--force originating from magnetic fields created by magnetospheric particle currents can also be neglected, in agreement with earlier conclusions of \cite{ertl}. Spontaneous pair creation still turns out to be insignificant, even within the very strong electromagnetic fields of polar regions.

In chapter 2 of what follows, we rediscuss  force--free surfaces associated with rotating magnets. Thereafter, in chapter 3, we display equations of {\em individual} as well as of {\em collective particle motion} in terms of a non-neutral two-component fluid description, and we then proceed to a description of appropriate tools for numerical treatment in chapter 4. Results will be given in chapter 5 and subsequently discussed in chapter 6.

\section{Vacuum Fields and Force--Free Surfaces}

The vacuum solution of Maxwell's equations for a homogeneously magnetized (ideally conducting) sphere, rotating with its vector of angular velocity $\bm{\o}$ inclined relative to its vector of magnetic dipol moment $\bm{\mu}$ by the angle $\chi$, as evaluated in \cite{deutsch}, called Deutsch--field, may be applied here in the near--field approximation. In addition, to account for the global electric charge of the rotating sphere, an electric monopol contribution $q_s$ is introduced. For Details about the Deutsch--field in the near--field approximation with a global electric charge of the rotating sphere we refer to \cite{wolf94}.

In the case of an ideally conducting sphere, the topography of the exterior field is known to be independent of the form of interior magnetization. An interior central magnetic point dipole, which may be chosen as an alternative model of magnetization, obviously would create the same electromagnetic vacuum field configuration. But the electric surface charge as well as the interior charge evoked by rotation clearly depend on the form of magnetization.

A central interior magnetic point dipole, for example, is consistent with the total interior electric charge $Q_i = \frac{2}{3}\frac{\mu}{r_L} \cos \chi $ and with the surface charge density $ \sigma = - \frac{\mu}{2 \pi r_L r_N^2}\, [ \cos \chi \cos^2 \vartheta + \sin \chi \cos \theta \sin \vartheta \cos (\p - \o t)]$, corresponding to the total surface charge $ Q_s = - \frac{2 \mu}{3 r_L} \cos \chi $ \footnote{Here we make use of two sets of spherical coordinates: one is referred to as the $\o$--system $(r, \vartheta, \p)$, where $r$ is the radial coordinate, $\vartheta$ is the angle measured against the rotation axis, and $\varphi$ is the angle relative to the plane spanned by the ${\bf x}_0$ and ${\bf y}_0$ axis, at rest in a chosen inertial frame of reference.

The other set of spherical coordinates is referred to as the $\mu$--system $(r, \psi, \l)$, in which $\psi$ is the angle relative to the magnetic dipol axis, and $\l$ is the angle against the plane spanned by the $\bm{ \mu}$ and $\bm{ \o}$ axis. As a consequence of these definitions, ${ \l = \p - \o [t - (r - r_N)/c] }$.}. 

Alternatively, for a homogeneously magnetized sphere, as considered throughout this paper, the surface charge density is $\sigma = - \frac{\mu}{4 \pi r_L r_N^2} [ \cos \chi ( 5 \cos^2 \vartheta- 3)+ 5 \sin \chi \cos \vartheta \sin \vartheta \cos(\p -\o t) ]$ and $ Q_i = \frac{4 \mu}{3 r_L} \cos \chi = - Q_s\, $.

The discontinuity of the tangential component of the magnetic surface field creates an electric surface current which results negligible under conditions given here.

\begin{figure}
\centering
\begin{minipage}{16cm}
\includegraphics[width=4.8cm, height=6.6cm]{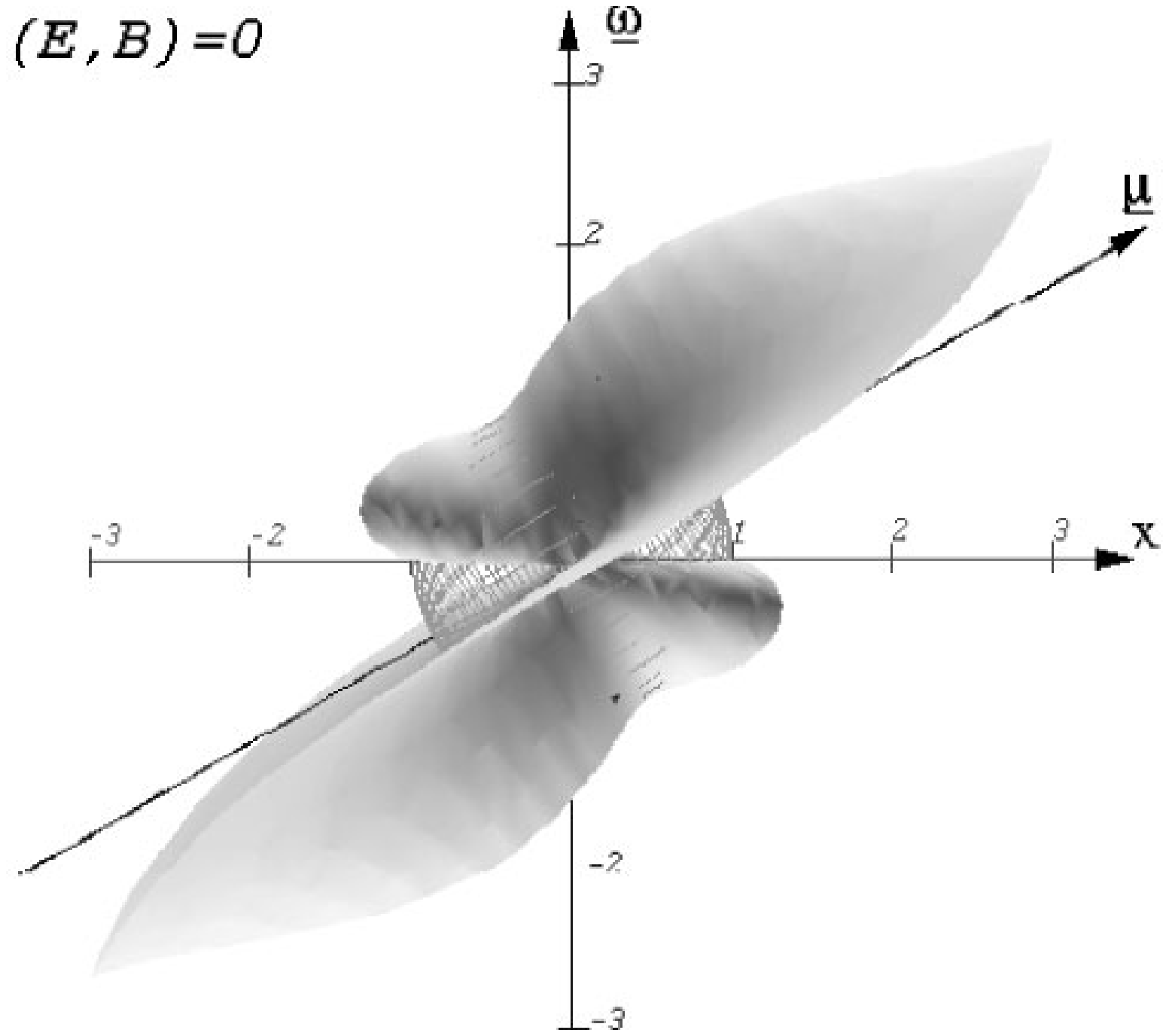}
\hspace{-0.3cm} 
\includegraphics[width=4.8cm, height=6.6cm]{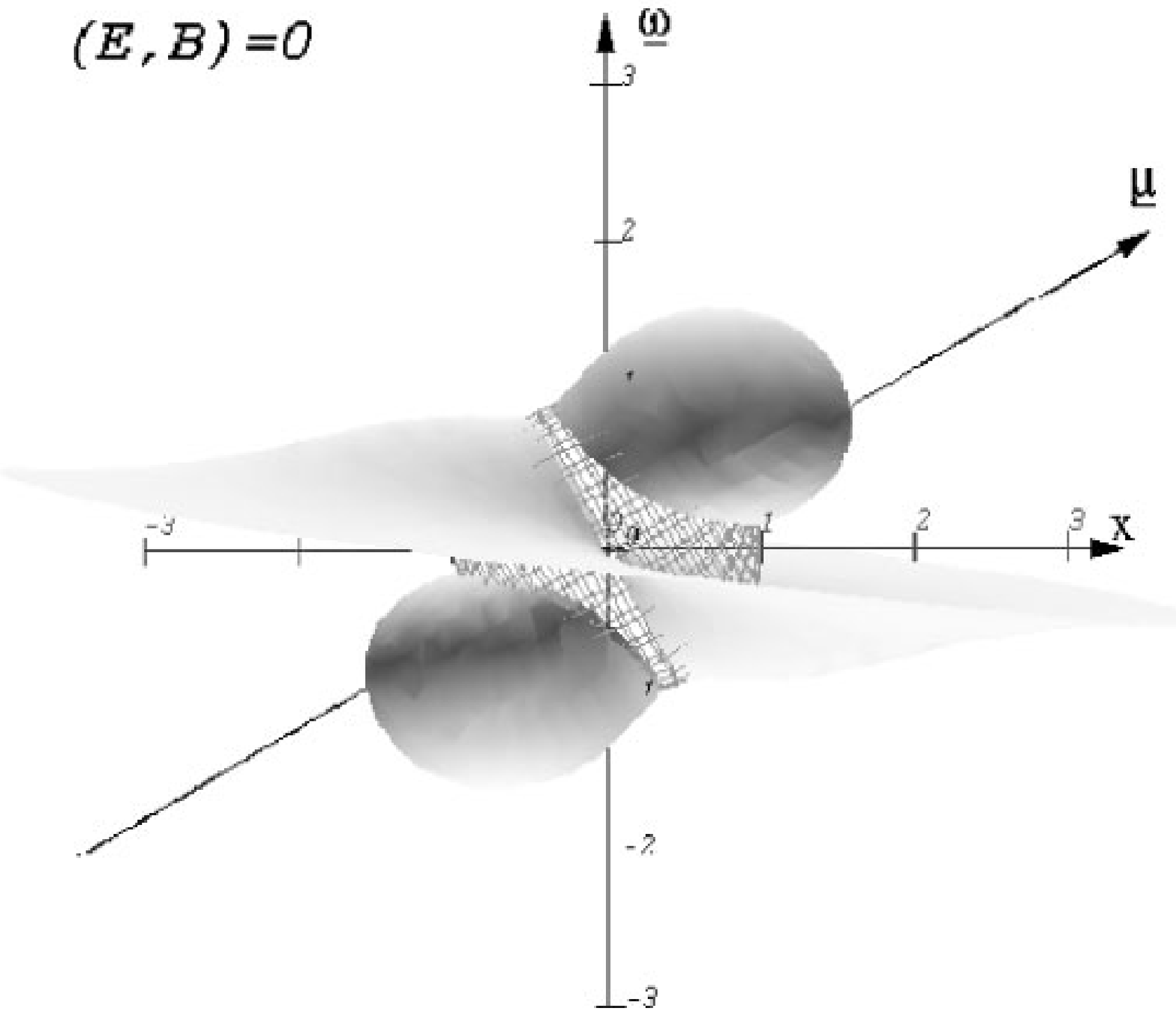}
\hspace{0.3cm}
\includegraphics[width=4.8cm, height=6.6cm]{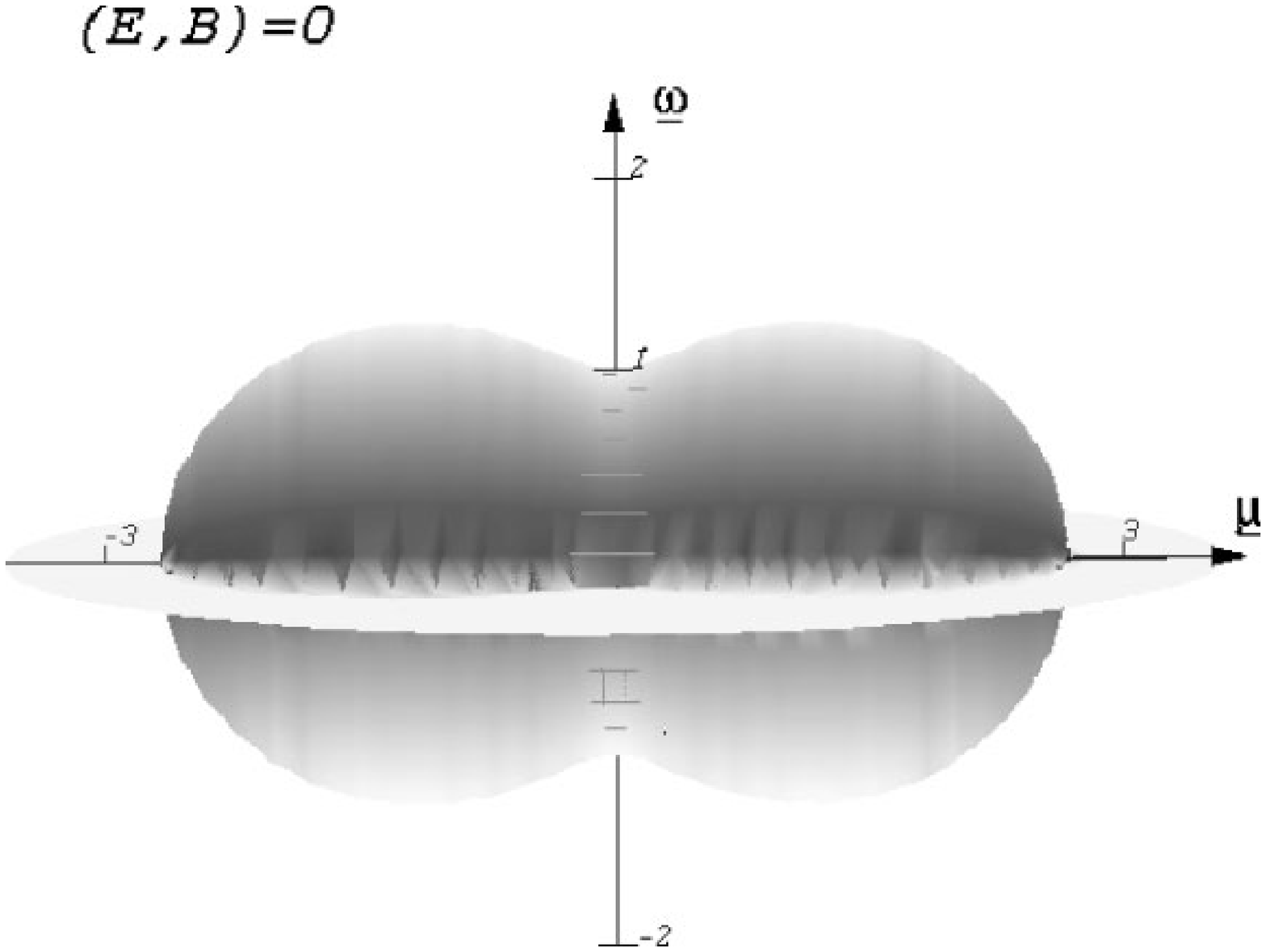}
\end{minipage}
\vspace{-1.5cm}
\caption{Force--Free Surface. Left: $\c = 60^o$, $q_s = 0 \,\,\mu/r_L$.
Middle: $\c = 60^o$, $q_s = 0.3 \,\,\mu/r_L$.
Right: $\c = 90^o$, $q_s = 0 \,\,\mu/r_L$. The unit of length
 on both axes: $r_N$. \label{ffs_60} }
\end{figure}

With the projection of the electric onto the magnetic vector, written in the $\mu$--system,
\begin{eqnarray} \nonumber
\frac{{\bf(E,B)}}{B} &=& -\frac{\mu \,k^3}{(k \,r)^4}\, 
\frac{1}{\sqrt{1 + 3 \cos^2 \psi}}\,\, 
 \left[ (r/r_N)^2  (\sin \chi \sin \psi \cos \l - 2\,q_s^{\prime} \cos \psi)  \right.
\\ \nonumber
&&  + 4 (\cos \chi \cos \psi
 - \sin \chi \sin \psi \cos \l) \cos^2
 \psi \left. - \sin \chi \sin \psi \cos \l \right] \
\end{eqnarray}
the FFS is given by
\begin{eqnarray}
\left(\frac{r}{r_N}\right)^2 &=&
 \frac{(\sin \chi \sin \psi \cos \l - 4 (\cos \chi \cos \psi 
\sin \chi \cos \l  \sin \psi ) \cos^2 \psi)}{(\sin \chi \sin \psi \cos \l 
 -  2 q_s^{\prime} \cos \psi)}\ ,
\end{eqnarray}
where $q_s^{\prime} := q_s \frac{r_L}{\m}$ is the dimensionless form of the total electric charge of the sphere. Some examples for the chape of the FSS are illustrated in figure \ref{ffs_60}.

\section{Non--Neutral Two--Component Plasma Fluid}

\subsection{Equations of Motion for the Plasma}

We consider a two--component ideal fluid, whereby each component (identified by the index $s = 1, 2$)consists of $N$ identical (classical) particles, electrons and protons in this case. For a relativistic macroscopic description of fluid motion in terms of the one--particle distribution function ${\cal F}_s(x^{\mu},p^{\mu})$ we start with the one--particle  Liouville--equation
\begin{equation}
\frac{d {\cal F}_s}{d \tau} = {\dot x}^{\mu} 
\frac{\partial {\cal F}_s}{\partial x^{\mu}}
+  {\dot p}^{\mu} \frac{\partial {\cal F}_s}{\partial p^{\mu}} = 0 \,,
\end{equation}
where $x^0 = ct$ is the time coordinate, $x^i$ are the spatial coordinates, $p^{\mu}$ are the four corresponding components of momentum \footnote{The signature of the metric tensor is (1,-1,-1,-1). Throughout this paper, Greek indices are running from 0 to 3, latin indices from 1 to 3. As stated before effects of general relativity (including gravitation) are left unaccounted for since, with the standard set of parameters the  Schwarzschild--radius is only about $r_s \approx 0.3 \, r_N$.}. Using the covariant form of  Hamilton's equations,
 $d x^{\mu}/d \tau = \partial {\cal H}/\partial p_{\mu}$ and 
 $d p^{\mu}/d \tau = - \partial {\cal H}/\partial x_{\mu}$, 
where ${\cal H}$ is the  Hamilton--function, one may write the covariant  Vlasov--equation in the form
\begin{equation}
\frac{\partial }{\partial x^{\mu}} ({\cal F}_s {\dot x}^{\mu}) + 
\frac{\partial }{\partial p^{\mu}} ({\cal F}_s {\dot p}^{\mu}) = 0 \, .
\end{equation}
With the definition of the  one--particle distribution function 
$ f_s( x^{\mu}, p^k)  := \int_{- \infty}^{\infty} \,
 {\cal F}_s(x^{\mu},p^{\nu}) \, d p^0 $ 
the  Vlasov--equation results (with the assumption ${\cal F} \rightarrow 0$ for $p^{\mu} \rightarrow \infty$) in:
\begin{equation}
\label{ko_vlasov_f}
\frac{\partial}{\partial x^{\mu}}(f_s \,{\dot x}^{\mu}) +
\frac{\partial}{\partial p^i}(f_s \,{\dot p}^i) = 0 \,.
\end{equation}
If $u^{\mu}=d x^{\mu}/d \tau$ are the four components of velocity, $a^{\mu}=d u^{\mu}/d \tau$ those of the acceleration, while $\gamma$ is the Lorentz--factor and $\tau$ the proper time, the  Vlasov--equation can be represented by
\begin{equation}
\label{vlasov}
\frac{\partial}{\partial t}(\gamma\,f_s) 
 +  c\,\frac{\partial}{\partial x^{i}} (u^{i}\,f_s)
 +  \frac{\partial}{\partial u^{i}}  (a^{i}\,f_s)= 0 \,.
\end{equation}   
The distribution function $f_s(x^{\mu},u^{i})$ delivers  by integration the average number density
\begin{equation}
n_s(x^{\mu}) = \int \,f_s (x^{\mu}, u^{i}) \, d^3 u \,
\end{equation}
and from there the average, macroscopic (four-component) velocity vector and the electric charge density is given by:
\begin{equation}
\bar{u}^{\mu}_s(x^{\mu}) = n_s^{-1} \,\int\, u^{\mu}\,f_s(x^{\mu},u^i)\, d^3 u \,\, , \,\,
\r(x^{\mu})= \sum_s \,e_s \,n_s(x^{\mu}).
\end{equation}
Likewise, we make use of the average, macroscopic (three-component) velocity vector ${\bar{v}_s}^i(x^{\mu})$ to define the electric current density
\begin{equation}
j^{i}(x^{\mu}) = \sum_s e_s  \int v^i f_s  d^3 u
  =  \sum_s e_s  {\bar{v}_s}^i(x^{\mu})
\end{equation}
constituting thereby the four components of the electric current density vector field $j^{\mu} = (c \r, j^i)$.

The latter, in consistency with (\ref{vlasov}), act as  source terms in  Maxwell's--equations $\partial_{\a} F^{\a \b} = \frac{4 \pi}{c} j^{\b}\, , \partial_{\a} F^{\star \,\a \b} = 0 $ , where $F^{\a\b}= \partial^{\a} A^{\b} -  \partial^{\b} A^{\a}$ is the electromagnetic field tensor and $F^{\star\,\a\b} = \frac{1}{2} \epsilon^{\a \b \gamma \delta}\,F_{\gamma \delta}$ its dual counterpart. $A^{\m}$ represents the electromagnetic potential and $\epsilon^{\a \b \gamma \delta}$ the Levi--Civita--tensor.

The average number density $n_s(x^{\mu})$ and velocity $\bar{u}^i_s(x^{\mu})$ fields of each of the two constituents of the plasma serve to describe macroscopic fluid motion with the help of the continuity equation
\begin{equation}
\frac{\partial n_s}{\partial t}
 + \frac{\partial }{ \partial x^i}( n_s {{\bar{v}}_s}^i) =
 - \int \frac{\partial}{ \partial u^{i}}(a^{i} f_s)\,d^3 u = 0 \,,
\end{equation}
obtained from (\ref{vlasov}) through integration over velocity space, adopting
${\vert {\bf a} \vert} f_s \rightarrow 0$ for ${\vert {\bf u} \vert} \rightarrow \infty $.
The energy--momentum equation
\begin{equation}
\label{impuls}
\frac{\partial \bar{u}_s^i}{\partial t}
 + \bar{v}_s^j \frac{\partial}{\partial x^j} \bar{u}_s^i = 
\frac{1}{(m_s n_s)} \left( - \partial_j p_s^{ij} + {\cal K}^i \right) \,,
\end{equation}
is obtained from (\ref{vlasov}) through multiplication with
$u^j$
and integration over the velocity space, where
\begin{equation}
p_s^{ij}(x^{\mu}) \!=\! \int m_s (u^i - \bar{u}_s^i) (u^j
 - \bar{u}_s^j) f_s d^3 u \!=\! m_s \!\!\int u^i u^j f_s d^3 u
 - m_s n_s \bar{u}_s^i \bar{u}_s^j
\end{equation}
is the pressure tensor and $m_s$ the rest mass of particles constituting the respective 
plasma component.

In what follows, we assume a collisionless, dispersionless (i.e.~cold), relativistic plasma. This assumption appears not implausible, since with a neutron star surface temperature of about $10^5 -10^7$ K and a particle density of about $n_s \approx 10^{12} \,{\rm cm}^{-3}$, as suggested by the  Goldreich \& Julian--model (1969), the plasma parameter $\Lambda  = ({4 \pi}/{3}) n_s \l_D^3 $ (where $\l_D = \sqrt{k_B T /(4 \pi n_s e^2)}$ is the  Debye--length) results in $\Lambda \approx 10^{7}$, so that $\Lambda \gg  1$ is given.

Thus, the right side of equation (\ref{impuls}) reduces to the (volume) force term
\begin{equation}
{\cal K}^{i} = m_s \int a^{i} f_s d^3 u \,,
\end{equation}
to which the  Lorentz--force is expected to deliver a major contribution ${{\cal K}^{\mu}}_{Lorentz} ={\eta_0} F^{{\mu} \nu} {\bar u}_{\nu},$ where $\eta_0 = e/(m c)$.

\subsection{Equations of Motion for Individual Electrically Charged Particles}

In addition to  the  Lorentz--force the classical equation of motion for an electrically charged {\em relativistic} particle subject to given ('external') electromagnetic fields has to account for radiation reaction forces. One such equation which frequently is called 'Lorentz--Dirac (LD) equation' \cite{dirac}(though  Abraham--Lorentz (AL) equation would be historically more correct) may be given the form
\begin{equation}
\label{equation of motion}
\frac{d u^{\mu}}{d \tau} = \eta_0 F^{\mu \nu} u_{\nu} 
+ \tau_0 G^{\mu \nu} u_{\nu}
\end{equation}
suggested by one of us \cite{th89}. Here,
\begin{equation}
\label{LD-tensor}
G^{\mu \nu} = G^{\mu \nu}_{LD} = \frac{1}{c^2} (u^{\nu} \frac{d^2 u^{\mu}}{d \tau^2} 
- u^{\mu}\frac{d^2 u^{\nu}}{d \tau^2})
\end{equation}
is the radiation force tensor and
$\tau_0 = 2 e^2/(3 m c^3)$
is the radiation constant. Unfortunately, this equation of motion exhibits serious deficiencies which have extensively been discussed in published literature and can be avoided through replacement of (\ref{LD-tensor}) by its first iteration often referred to as 'Lorentz--Dirac--Landau (LDL) equation' (due to its extensive discussion in \cite{landau} well known textbook). In this approximation \footnote{One of us \cite{th94} has argued that in a quantum-mechanical frame self-consistency of classical electrodynamics suggests (\ref{LDL-tensor}) to be the correct form of the radiadion tensor $G^{\mu \nu}$}, the radiation tensor may be written
\begin{equation}
\label{LDL-tensor}
G^{\mu \nu} = \eta_0 u_{\lambda} \partial^{\lambda} F^{\mu \nu} +
\frac{1}{c^2} (u^{\mu}_{LL} u^{\nu} - u^{\mu} u^{\nu}_{LL} ),
\end{equation}
with $u^{\mu}_{LL} = \eta_0^2 F^{\mu \nu}F_{\nu \lambda} u^{\lambda}$. In our numerical work presented further below we adopt that {\em locally} (i.e.~within appropriately small intervals of space and time coordinates) the electromagnetic field is (approximately) homogenous in space and constant in time, in which case the radiation tensor (\ref{LDL-tensor}) further reduces to
\begin{equation}
\label{LDL_constant}
G^{\mu \nu} = G^{\mu \nu}_{const} = \tau_0  \eta_0^2  
\left[ F^{\a}_{\,\,\,\nu} F^\nu_{\,\,\,\lambda} u^\lambda +
F^{\l}_{\,\,\,\varrho} F_{\l}^{\,\,\,\p} 
u^{\varrho} u_{\p}\, u^{\a} \right] . 
\end{equation}
Under these premises individual particles of each of the two plasma components, according to s = 1 or 2, inside the proper interval of space and time coordinates (i.e.~inside the corresponding 'volume element')  are subject to the {\em same} electromagnetic forces. Consequently, the (macroscopic field) equations for the plasma fluid 
\footnote{Density effects on radiation and radiation reaction remain unaccounted for. Remarkably, since in (\ref{impuls2}) the pressure term is absent, the number density does not appear. Clearly, (\ref{impuls2}) governs the macroscopic velocity fields $\bar{u}_s^\lambda$.}
immediately follows from (\ref{impuls}) and (\ref{equation of motion}) with (\ref{LDL_constant})
\begin{equation}
\label{impuls2}
 \bar{u}_{\beta,s} \partial^{\beta} \bar{u}_s^{\alpha} =
\eta_0 F^{\alpha \beta} \bar{u}_{\beta,s}
+  \tau_0  \eta_0^2  
\left[ F^{\a}_{\,\,\,\nu} F^\nu_{\,\,\,\lambda} \bar{u}_s^\lambda +
F^{\l}_{\,\,\,\varrho} F_{\l}^{\,\,\,\p} 
\bar{u}_s^{\varrho} \bar{u}_{\p,s} \, \bar{u}_s^{\a} \right] . 
\end{equation}

\subsection{Exact Solutions of the Equation of Motion for Individual Particles in Homogenous and Constant Fields}

In what follows we make use of exact analytical solutions of the equation of motion (\ref{LDL_constant}) for individual electrically charged particles in locally constant and homogenous electromagnetic fields \cite{th94}. Given a coordinate system with 
${\bf e}_3 = {\bf B}$, ${\bf e}_2 = [{\bf E,B}]$ and 
${\bf e}_1 = [{\bf e}_2,{\bf e}_3 ]$
and excluding null fields \footnote{Inside the magnetosphere,  $|{\bf E}| \ll |{\bf B}|$} (i.e.~fields with simultaneously vanishing  Lorentz--invariants  $({\bf E,B})$ and ${\bf E}^2 - {\bf B}^2$) and restricting further to $E_1 \ne 0$ and $B \ne 0$, the solution of (\ref{LDL_constant}) is given by:
\begin{eqnarray}\nonumber
 u^{\a}(\tau) &=& 
 \gamma \,a(\tau) \left\{  
C_0 \left(
\begin{array} {r@{}} 
\mbox{cosh} \l \tau \\ \b \frac{\l}{\Omega_L} \mbox{sinh} \l \tau\\ 
- \b \mbox{cosh} \l \tau \\\frac{\omega}{\Omega_L} \mbox{sinh} \l \tau 
\end{array}\right)
+ C_3 \left( \begin{array} {r@{}} 
\mbox{sinh} \l \tau \\ \b \frac{\l}{\Omega_L} \mbox{cosh} \l \tau \\ 
- \b \mbox{sinh} \l \tau \\  \frac{\omega}{\Omega_L} \mbox{cosh} \l \tau
\end{array}\right)
\right\} \\ 
&+& \gamma \, b(\tau)  \left\{ 
C_1 \left(
 \begin{array} {r@{}} 
\b \sin \omega \tau \\\frac{\omega}{\Omega_L} \cos \omega \tau \\ 
- \sin \omega \tau \\  - \b \frac{\l}{\Omega_L} \cos \omega \tau
\end{array}\right)
+ C_2 \left(
\begin{array} {r@{}} 
- \b \cos \omega \tau \\ \frac{\omega}{\Omega_L} \sin \omega \tau \\ 
\cos \omega \tau  \\ -  \b \frac{\l}{\Omega_L} \sin \omega \tau 
\end{array}\right)
\right\}
\end{eqnarray}
with 
$\l = {\rm sign}(E_3) \frac{e}{m c} \sqrt{- \frac{1}{2} ( B^2 - E^2) + 
\frac{1}{2} 
\sqrt{(B^2 - E^2)^2 + 4\, ({\bf E,B})^2 }}$, $ \Omega_L = \frac{e}{m c} B $,
$ \b = \frac{ E^2 + B^2 - \sqrt{(B^2 - E^2)^2  
+ 4 \, ({\bf E ,B})^2}}{ 2\,E_1\,B}$,
$ \omega  = \frac{e}{m c} \sqrt{ \frac{1}{2} ( B^2 - E^2) + 
\frac{1}{2} 
\sqrt{(B^2 - E^2)^2 + 4 ({\bf E,B})^2 }}$, 
 $ \gamma = \frac{1}{\sqrt{2}} \sqrt{\frac{B^2 + E^2}{\sqrt{(B^2 - E^2)^2 + 
4\, ({\bf E,B})^2 }}  + 1}$ , 
\begin{eqnarray}
\left(\begin{array} {r@{}} C_0 \\ C_1 \\ C_2 \\ C_3  \end{array}\right) &=&
\gamma \left(\begin{array} {r@{}} 
u^0(0) + \b u^2(0) \\ \nonumber
\frac{\omega}{\Omega_L}u^1(0) - \b\frac{\l}{\Omega_L} u^3(0) \\ \nonumber
\b u^0(0) + u^2(0)   \\ \nonumber
\b \frac{\l}{\Omega_L}u^1(0) + \frac{\omega}{\Omega_L} u^3(0) \\\nonumber
\end{array}\right) \, 
\end{eqnarray}
and the radiation parameters $a(\tau) = \left[ 
(C_0^2 - C_3^2) - (C_1^2 + C_2^2)\exp[-2 \tau_0 (\l^2 +
\omega^2) \tau ] \right]^{-\frac{1}{2}} $ ,
$ b(\tau) = a(\tau) \exp[- \tau_0 (\l^2 + \omega^2) \,\tau ]$, where $C_{\m} C^{\m} = 1\,$ .

If the electric field is parallel (or antiparallel)to the magnetic one ($E_1 =0$) this solution reduces to:
\begin{eqnarray} 
u^{\a}(\tau) &=& 
a(\tau)   \left\{   \nonumber
u^0(0) \left(
\begin{array} {l@{}} 
\mbox{cosh} \l \tau \\ 0\\ 0 \\  \mbox{sinh} \l \tau 
\end{array} \right)
+ u^1(0) \left( \begin{array} {l@{}} 
0\\ \mbox{cos} \omega \tau \\ 
- \mbox{sin} \omega \tau \\ 0
\end{array}\right)
\right\} \\ 
&+&\, b(\tau)   \left\{ 
u^2(0) \left( \begin{array} {l@{}} 
0\\ \sin \omega \tau \\ 
\cos \omega \tau \\ 0
\end{array}\right)
+ u^3(0)\left(
\begin{array} {l@{}} 
\sinh \l \tau \\ 0 \\ 0 \\\cosh \l \tau 
\end{array}\right)
\right\}\,.
\end{eqnarray}

The dynamics of a charged particle starting at rest is characterized by the following properties: Within a regime of small proper time, $\tau \ll 1 / {\tau_0 (\omega^2 + \l^2)}$, the radiation parameters are about $a(\tau) \approx b(\tau) \approx 1$, while particle motion is a composit of gyration around magnetic field lines, $[{\bf E,B}]$--drift, and acceleration along magnetic field lines, due to the projection of the electric field vector onto to the tangent to the magnetic field line.

Within a regime of large proper time, $\tau \rightarrow \infty$, gyrations around magnetic field lines are damped corresponding to $b(\tau) \rightarrow 0$, while acceleration along magnetic field lines proceeds correspondig to $a(\tau) \rightarrow 1$, superimposed by $[{\bf E,B}]$--drift and curvature drift.

In the magnetospherical regime considered here ($|{\bf E}| \ll |{\bf B}|$), where $\omega \approx \Omega_L, \l \approx \Omega_L \frac{E}{B} \sin \a$ and
\begin{eqnarray}
\frac{1}{\tau_0 (\omega^2 + \l^2)} \approx \left\{
\begin{array}{l@{\quad  \quad}l}
5\cdot 10^{-16} \,{\rm s}& {\rm for \,\,electrons}\\
3\cdot 10^{-6} \,{\rm s}& {\rm for \,\,protons\,\, ,}
\end{array}
\right.
\end{eqnarray}
gyrations around magnetic field lines are damped.

\section{Numerical Procedures on the Grid}

\subsection{Velocity Components}

Evolution in time of the four components of the velocity vector $u^{\a}(t)$ of individual electrically charged particles, from $u^{\a}(t_0)$ to $u^{\a}(t_1)$  with $\Delta t = t_1 - t_0$ (with substitution of proper time $\tau$ by the time coordinate $t$, involving the integration of $u^0(\tau)$ over $\tau$) according to what was discussed before, is governed by the equation of motion (\ref{equation of motion}) with (\ref{LDL-tensor}) (within appropriately small intervals of space and time coordinates). The same holds for a plasma velocity field, in which, according to the dispersion-free fluid model used here, individual particles represent co-moving fluid elements,
\begin{equation}
\label{integration2}
\fl u^{\a}(\tau_1) - u^{\a}(\tau_0) = 
u^{\a}(t_1) - u^{\a}(t_0) 
= \int_{t_0}^{t_1} \frac{d u^{\a}(t)}{d t} d t 
=
 \int_{t_0}^{t_1} \left(\partial_t+ v_j(t,{\bf x}) \partial^j \right)
u^{\a}(t,{\bf x}) \,d t \, .
\end{equation}
The first term in the last integral of (\ref{integration2}) refers to the {\em explicit} time dependence of the velocity field (i.e.~the change with time at a fixed point {\bf x} in space), whereas the second term describes its {\em implicit} time dependence (i.e.~the additional change with time  of a co-mpoving particular fluid element). 

Here we are interested in the change in time of the velocity field at a fixed point {\bf x} in configuration space. Due to the given initial condition $u^{\a}(\tau_0 = 0) \equiv u^{\a}(t_0,{\bf x})$ in each iteration step (where $u^{\a} (t,{\bf x})$ denotes the velocity at a given point in configuration space) it follows:
\begin{equation}
\label{integration3}
u^{\a} (t_1,{\bf x}) = u^{\a}(t_1) - \int_{t_0}^{t_1} 
v_j(t,{\bf x}) \partial^j  \,u^{\a}(t,{\bf x})\, d t\,.
\end{equation}

Velocity components are required on grid points of a spherical grid, where the components of the electric field vector are suggested to be known from electric charge density averaged over each individual cell.

The integral in (\ref{integration3}) can be evaluated in a first order scheme in time with
$v := -(t_1 -t) \Rightarrow \partial_t v = 1$,
\begin{equation}
\label{diff_int}
\fl u^{\a} (t_1,{\bf x}) = u^{\a}(t_1) -  
(t_1 -t_0)v_j(t_0,{\bf x}) \partial^j \,u^{\a}(t_0,{\bf x})
- \int_{t_0}^{t_1} 
(t_1 -t) \,\partial_t [v_j(t,{\bf x}) \partial^j u^{\a}(t,{\bf x})]\, d t\,
\end{equation}
by neglecting the integral in this evaluation. Differentiation with respect to spatial coordinates is performed numerically in a second order centered difference scheme (boundaries are treated separately, also in second order). The numerical scheme results in
\begin{eqnarray}
&&u^{\a} (t_1,{\bf x}_i) \, = \,u^{\a}(t_1) - \\ \nonumber
&&\frac{\Delta t}{\gamma} \left(
v_{r,i} \frac{1}{\Delta r} + v_{\t,i} \frac{1}{r \Delta \t}
+ v_{\p,i} \frac{1}{ r \sin \t \Delta \p}\right) \frac{1}{2}\,[u^{\a}(t_0,{\bf x}_{i+1}) - u^{\a}(t_0,{\bf x}_{i-1})] \, ,
\end{eqnarray}
where the index $i$ now characterizes the $i^{th}$ grid point (e.g.~with respect to ${\bf e}_r$--direction, etc.).

\subsection{Electromagnetic Field Components}
\label{em}

Numerical integration of electromagnetic field equations is performed applying a scheme developped in our group by \cite{laue}, implying a (complete and orthonormal) system of spherical vector harmonics ${\bf P_{n m}, B_{n m }, C_{n m}}$ (e.g.~\cite{morse}),
\begin{eqnarray}\nonumber
{\bf P}_{n m}(\t, \p)&=& \phantom{+}\, {\bf e}_r X_{n m}\,\,,
{\bf B}_{n m}(\t, \p)= [{\bf e}_r,{\bf C}_{n m}(\t, \p)]
= \frac{r}{\sqrt{n(n+1)}}\, (\Bn, X_{n m}) \,,\\ \nonumber
{\bf C}_{n m}(\t, \p)&=& - \,[ {\bf e}_r, {\bf B}_{n m}(\t, \p)] 
=\frac{r}{\sqrt{n(n+1)}}\,[\Bn, {\bf r} X_{n m}]\,, 
\end{eqnarray}
where $n \in \{0,1,2,...\}$, $m \in \{-n,...,0,...,n\}$ and the spherical harmonics  $X_{n m}(\t,\p)$ are defined with the help of the associated  Le\-gen\-dre--function
$ X_{n m}(\t,\p) = e^{i m \p} \, P_n^m (\cos \t)$ permitting the expansion of the elctromagnetic vector potential
\begin{equation}
\fl  {\bf A}(r,\t,\p,t) =
\sum_{n, m}\, [\, p_{n m}(r,t) {\bf P}_{n m}(\t,\p)
+ \,b_{n m}(r,t){\bf B}_{n m}(\t,\p)
+  c_{n m}(r,t){\bf C}_{n m}(\t,\p) \,] \,.
\end{equation}
For example, the vector potential of the  Deutsch--field can be represented in the form
\begin{eqnarray}\nonumber
{\bf A}(r,\t,\p) &=& t\, \frac{r_N^2}{r^4} \cos \chi \,{\bf P}_{2 0} 
- i e^{-i t} \frac{h_2(r)}{H_2(r_N)} \sin \chi \,{\bf P}_{2 1}\\ \nonumber
&+& t  \frac{r_N^2}{r^4} \cos \chi \, {\bf B}_{2 0} 
- i e^{-i t} \frac{H_2(r)}{r^2 \, H_2(r_N)} \sin \chi \,
{\bf B}_{2 1}\\ 
&+&  \frac{1}{r^2}  \cos \chi  \,{\bf C}_{1 0} 
+ e^{-i t} \frac{h_1(r)}{r_N^2 \,h_1(r_N)}
 \sin \chi \,{\bf C}_{1 1},
\end{eqnarray}
with $h_1(r)= -e^{i r}\frac{(1+i/r)}{r} \,,\, h_2(r)=i\,e^{i r}\frac{(1+ 3 i/r^2 - 3/r^2)}{r^2}\,,\, H_2(r)= e^{i r}\frac{(6r - r^3)+ i(6 - 3 r^2)}{r^2}\,$.

From there the magnetic field vector is calculated from ${\bf B} = [\Bn,{\bf A}]$. Exploiting the gauge invariance of the four component vector potential to eleminate $A^0$ the electric field vector is calculated by ${\bf E} = - \partial_t {\bf A}$. 

To evaluate the total electromagnetic field in terms of its expansion coefficients,
we add the expansion coefficients of the Deutsch (vacuum) field to the expansion coefficients of the the plasma (different gauges are used for the two components). The electric (scalar) potential $A_0(r,\psi,\alpha)$ is determined from the charge density  $\varrho(r,\t,\p)$:  $A_0( {\bf r} ) = \int \frac{ \varrho( {\bf r}^{\prime})}
{|{\bf {r-r}}^{\prime} |} d^3 r^{\prime}\,$, assuming $A_0(r_N)=A_0(\infty)$.
Furthermore, $\frac{1}{|{\bf {r-r}}^{\prime} |} = \sum_{n,m} w_{n m}
\frac{r_<^n}{r_>^{n+1}}\, X_{n m}^{\ast}(\t^{\prime},\p^{\prime})\, 
X_{n m}(\t,\p)$,
where $w_{nm} = {(n-m)!}/{(n+m)!}$
and, by definition, $r_<$ ($r_>$) refers to the lower (upper) 
limit of the considered range
of $|{\bf {r}}|$ and $|{\bf {r^{\prime}}}|$, respectively,
so that
\begin{equation}
\fl A_0( {\bf r} ) = \sum_{n,m} X_{n m}(\t,\p) \int w_{n m}
\frac{r_<^n}{r_>^{n+1}} \varrho( {\bf r}^{\prime}) X_{n m}^{\ast}(\t^{\prime},
\p^{\prime})   d^3 r^{\prime}
=   \sum_{n,m} x_{n m}(r) X_{n m}(\t,\p)
\end{equation}
with
$ x_{n m}(r) = \int w_{n m}
\frac{r_<^n}{r_>^{n+1}} \varrho( {\bf r}^{\prime}) X_{n m}^{\ast}(\t^{\prime},
\p^{\prime}) d^3 r^{\prime} 
= \int \frac{r_<^n}{r_>^{n+1}}\, \varrho_{n m}(r^{\prime})\,
r^{\prime \,2}\, d r^{\prime} $\,. 
Here
$ \varrho_{n m}(r^{\prime}) = \int \int  w_{n m} \varrho({\bf r}^{\prime}) 
X_{n m}^{\ast}(\t^{\prime},\p^{\prime}) \sin (\t^{\prime}) 
d \vartheta^{\prime} d \varphi^{\prime}$\,
are the expansion coefficients of charge density.

For a thin spherical shell of thickness $\Delta r$ these expansion coefficients are 
\begin{equation}
x_{n m}(r) \approx \sum_j \Delta r \, r_j^2 \, \frac{r_<^n}{r_>^{n+1}}\, 
\varrho_{n m}(r_j) = \sum_j \frac{r_<^n}{r_>^{n+1}}\, q_{n m}(r_j)
\end{equation}
with $q_{n m}(r_j) = \Delta r\, r_j^2 \,\varrho_{n m}(r_j)$. 

The modes of image charges can be seen as additional modes of the surface charge density. The image charge at position  ${\bf r}_1$ is characterized by
$r_{sp}=r_N^2/r_1$ and $Q(r_{sp})=-(r_N/r_1)Q(r_1)$.
In general,
${\bf r_{sp} = r_<}$
so that $q_{n m}(r_{sp}) = - \frac{r_N}{r_1}q_{n m}(r_1), 
x_{n m}(r_{sp}) = \frac{r_N^n}{r_>^{n+1}}{\mbox{\=q}}_{n m}(r_N)$, with ${\mbox{\=q}}_{n m}(r_N)= -\frac{r_N^n}{r_1^{n+1}}\, q_{n m}(r_1)$.

The total potential at grid points ${\bf r}=(r_i,\t_j,\p_k)$ inside the spherical volume is written ${\cal A}_{n m}(r_i) = \sum_{j<i} r_j^n q_{n m}(r_j)$ 
and at grid points outside the spherical volume ${\cal B}_{n m}(r_i) = \sum_{j>i} \frac{1}{r_j^{n+1}} q_{n m}(r_j)$.  From that, the potential is
\begin{eqnarray}\nonumber
\label{A0}
A_0(r_i,\p_j,\t_k) &=&\sum_{n,m} {\cal D}_{n m}(r_i) \,X_{n m}(\t_j,\p_k) \, ,
\end{eqnarray}
with ${\cal D}_{n m}
 = {\cal A}_{n m}\,r_i^{-(n+1)} + {\cal B}_{n m}(r_i)\, r_i^n $,
resulting in the coefficients of the electric field 
$p_{m n}(r) = - \partial_r {\cal D}_{n m}(r)$, 
$ b_{m n}(r) = - \frac{\sqrt{n(n+1)}}{r} {\cal D}_{n m}(r)$, 
$c_{m n}(r) = 0 $.

\subsection{Continuity Equation}

In order to integrate the continuity equation $\partial_{\m}j^{\m} =0$ the  method of {\it flux corrected transport} (FCT) is used. This method bases on an algorithm developped by \cite{boris, book, zalesek}. In a first step a numerical 'low-order' scheme is applied introducing sufficient local numerical diffusion in order the get the numerical integration of the transport equation stable and monotonous. In a second 'high-order' step the introduced numerical diffusion is eliminated as far as possible. For details regarding FCT we refer to \cite{zalesek}. 

We extended this procedure to three dimensions, which is shown in \ref{FCT}. At the end  we get a stable conservative discretisation scheme with low numerical diffusion to integrate the continuity equation numerically.

\subsection{Particle Injection}

Of the three frequently discussed injection mechanisms -- (1) emission from the spherical surface (as determined by the electric field topography at the surface and surface charge density), (2) invasion of particles from outer regions, and (3) electron--positron pair creation from photon decay -- we need to consider only (1) here. The rate of particle injection from the surface is chosen to be proportional to the magnitude of the electric field component projected onto the tangent to the corresponding magnetic field line $E_{||}= \rm{sign}(\cos \psi)({\bf E,B})/B$, if the sign of surface charge density agrees with the sign of $E_{||}$ at the respective point on the surface.

\subsection{Reproducing the Magnetospheric Configuration}

All simulations carried out in order to get the results presented in this paper are started from the vacuum case. For $\chi=0$ ($\chi=\pi$) the electric field in vacuum allows emission of electrons (protons) only, while emission regions of electrons and protons are equally large for $\chi = \frac{\pi}{2}$. For $0 < \chi < \frac{\pi}{2}$ the emission of electrons and for $\frac{\pi}{2} < \chi < \pi$ the emission of protons predominates. 

We study the magnetosphere of an initially non--charged, homogenous magnetized sphere up to $20\,r_N$ with the standard set of parameters. The following grid sizes are used in the numerical simulation:
\begin{eqnarray*}
&&\Delta t^{\prime} =  5.0 \cdot 10^{-5}\,\,\,\,\mbox{if}\,\, \chi=0,\chi=\pi \,\, {\rm and} \,\,
\Delta t^{\prime} =  2.5 \cdot 10^{-5} \,\,\,\,\mbox{if}\,\, 0<\chi < \pi \\
&& \Delta r^{\prime} = 3.99 \cdot 10^{-4}\\
&&\Delta \t^{\prime}= \Delta \p^{\prime} =  2.06 \cdot 10^{-4} \,\,\,\,ib   \,\, {\rm and} \,\,
 \Delta \t^{\prime} = \Delta \p^{\prime} =  4.12 \cdot 10^{-3}
 \,\,\,\,ob\, ,
\end{eqnarray*}
where $ib$ destigates the inner border and $ob$ the outer border. This resolution implies that e.g.~for the (anti)parallel rotator a fluid element can cross the radial simulation extension 50 times during simulation time, which corresponds to a $114.6^{\circ}$ rotation of the rotating oject. For the oblique rotator, the simulation time corresponds to a rotation of $57.3^{\circ}$, so that a fluid element can cross the radial simulation extension 25 times. Consequently, the  simulation time is large enough to reproduce existing quasi-stationary magnetospheres. We study the rotator exemplary for several inclination angles: $\chi=0^{\circ},30^{\circ},45^{\circ},60^{\circ},75^{\circ},90^{\circ}, 120^{\circ}$.

\section{Results} 

We present and discuss our results under the aspect whether the formation of quasi-stationary magnetospheres can be verified and if so, what can be said of its structural features. In this context we investigate the verification of the predictions of the  Goldreich \& Julian--model and other authors in the special case of the aligned rotator. In the case of the inclined and orthogonal rotators (in which little is known from published literature) we investigate the structure of the quasi-stationary magnetospheres as well and stress in general the question concerning the typical particle number densities, currents and average particle energies inside the magnetospheres in the regime of the FFS.

\begin{figure}[h]
\centering
\begin{minipage}{16.5cm}
\includegraphics[width=8cm, height=5cm]{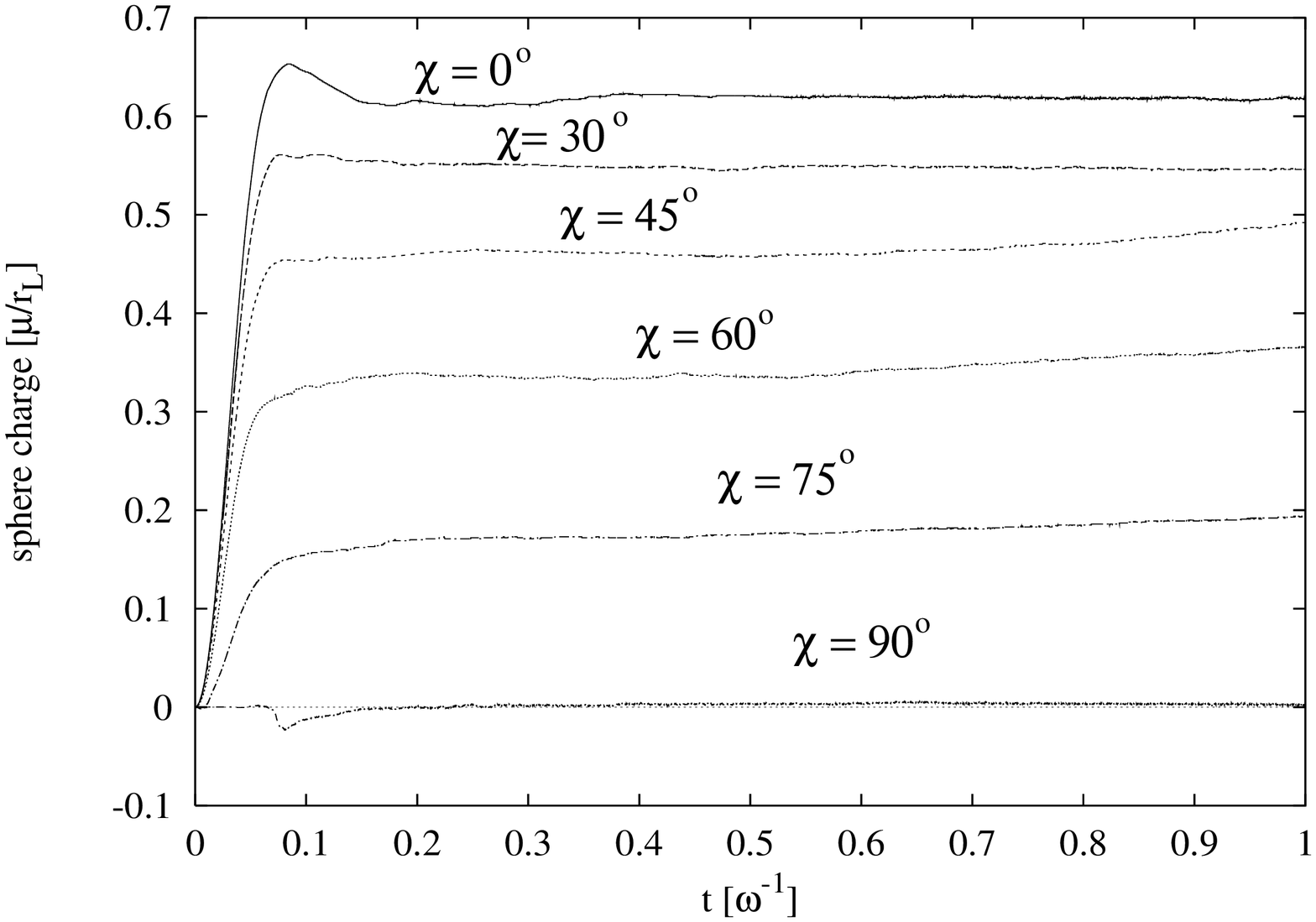}
\includegraphics[width=7cm, height=4.8cm]{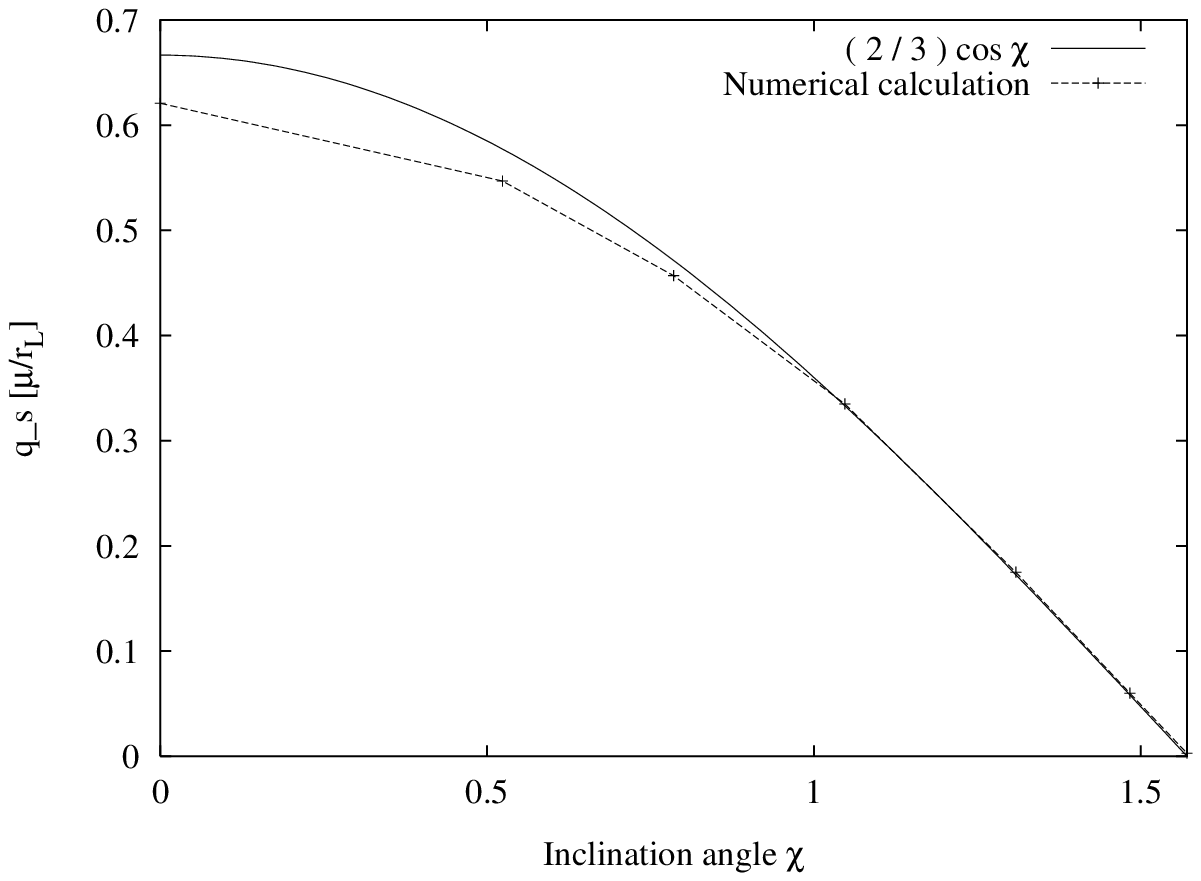}
\end{minipage}
\caption{Left fig.: Sphere charge depending on the time for different inclination
  angle. Right fig.: Sphere charge depending on the inclination angle in the case of quasi-stationary 
magnetospheres. \label{fig_qoo}}
\end{figure}

Figure \ref{fig_qoo} (left) demonstrates the development of the total electric charge of the sphere (over time) for various values of the inclination angle $\chi$, ranging from $\chi = 0$ to  $\chi = \pi/2$. Obviously, an asymptotic value is reached in each of these cases within time intervals well below $t \approx 0.2 \o^{-1}$, corresponding to a rotation of $\approx 11.2^{\circ}$. This behaviour can be seen as a clear indication for the rapid formation  of quasi-stationary magnetospheric configurations for all inclination angles. 

Furthermore, a prametrization of the electric monopole of the rotating sphere after building quasi-stationary magnetospheres in term of the inclination angle is possible and leads to (figure \ref{fig_qoo}, right): $q_s = \frac{2}{3}\frac{\mu}{r_L} \cos \chi $.

\begin{figure}[h]
\centering
\begin{minipage}{14.4cm}
\includegraphics[width=7.2cm]{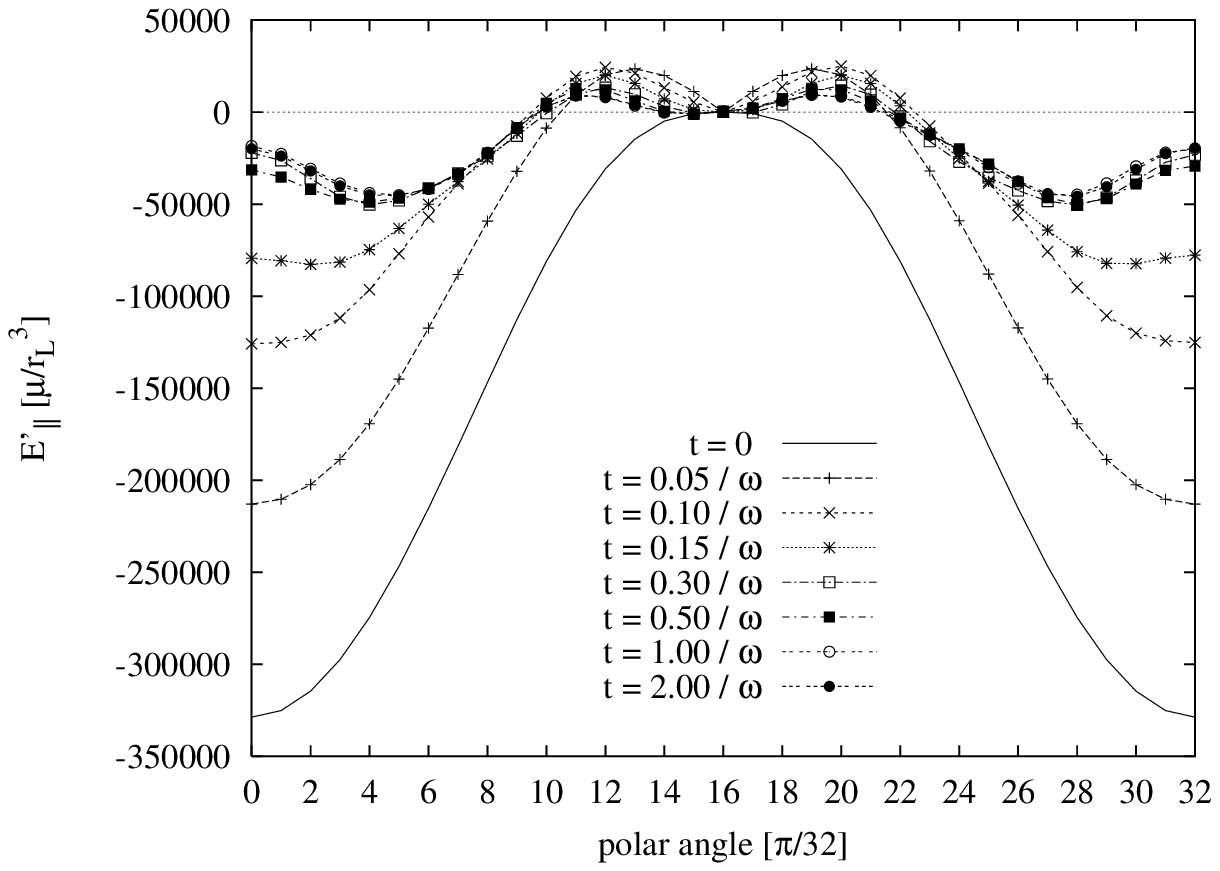}
\includegraphics[width=7.2cm]{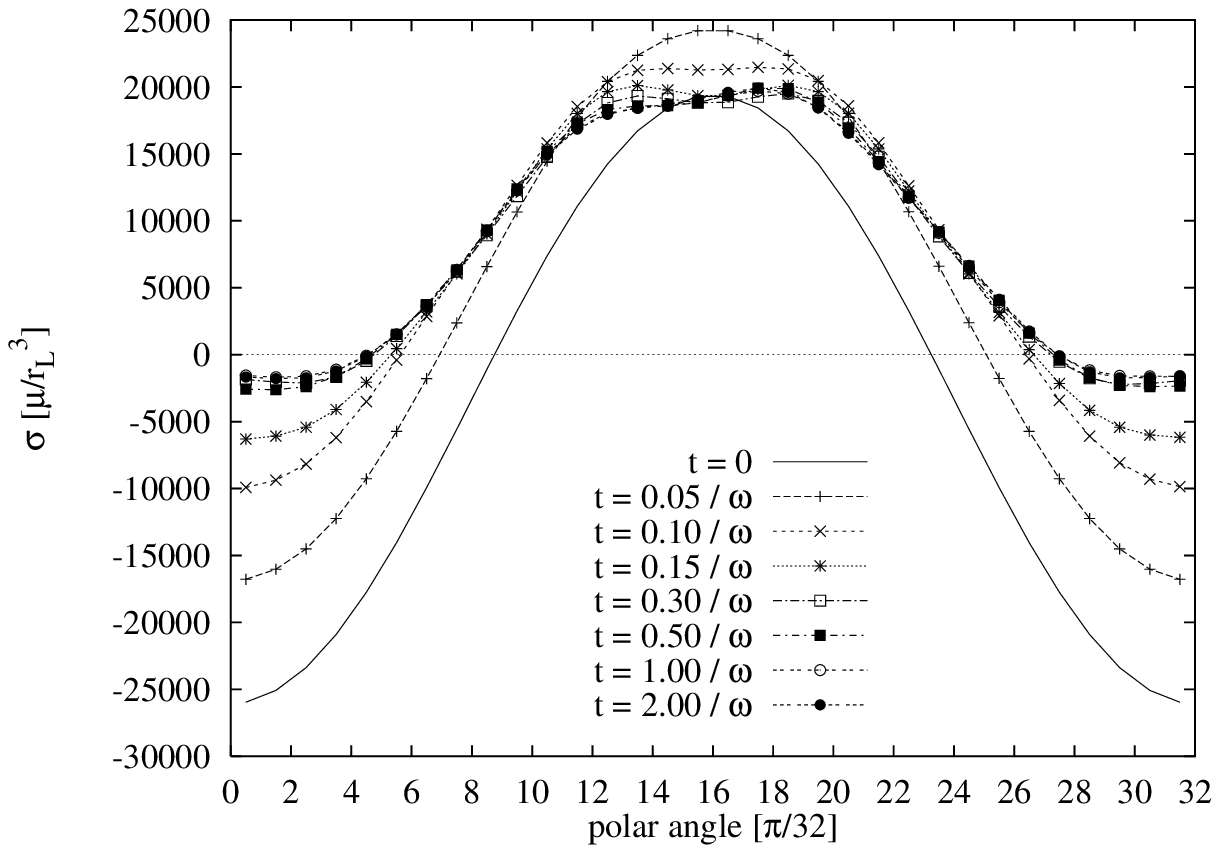}
\hfill
\includegraphics[width=7.2cm]{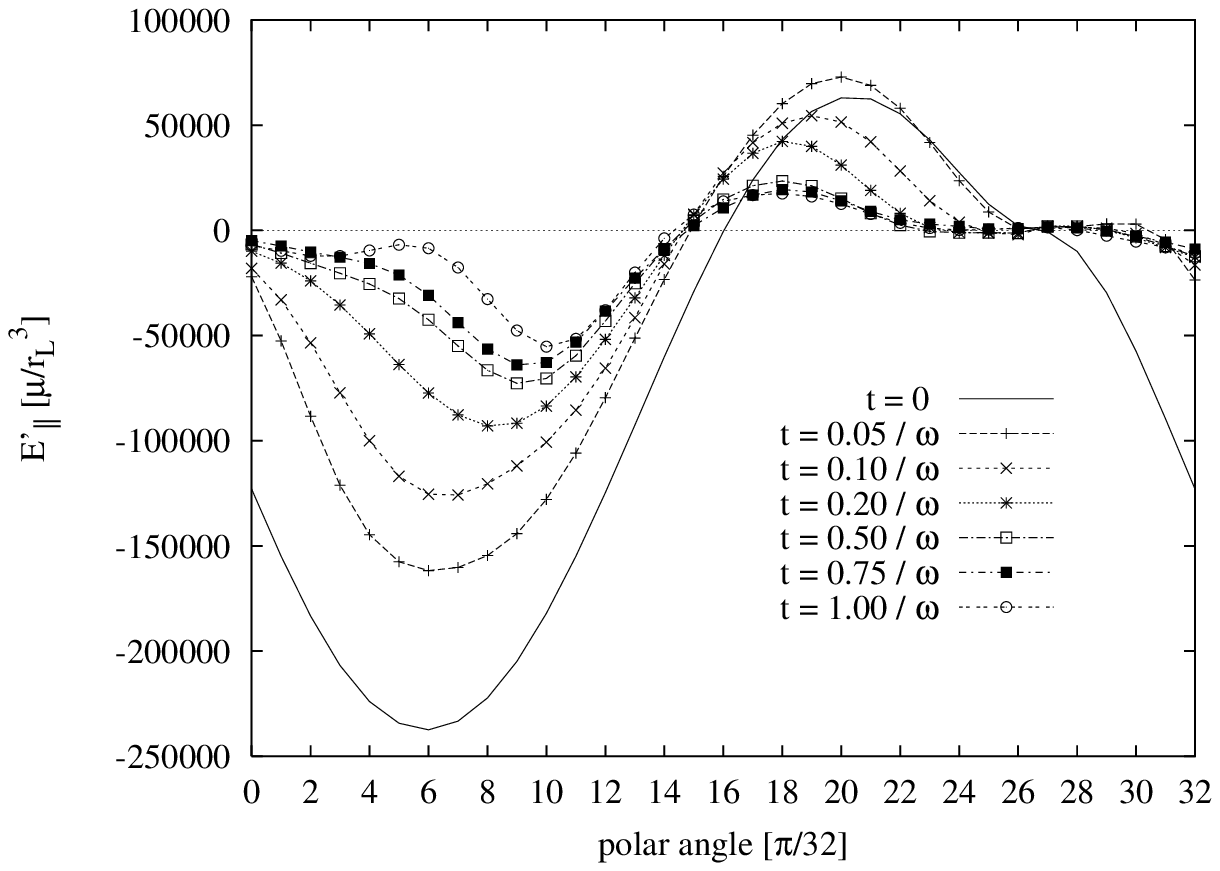}
\includegraphics[width=7.2cm]{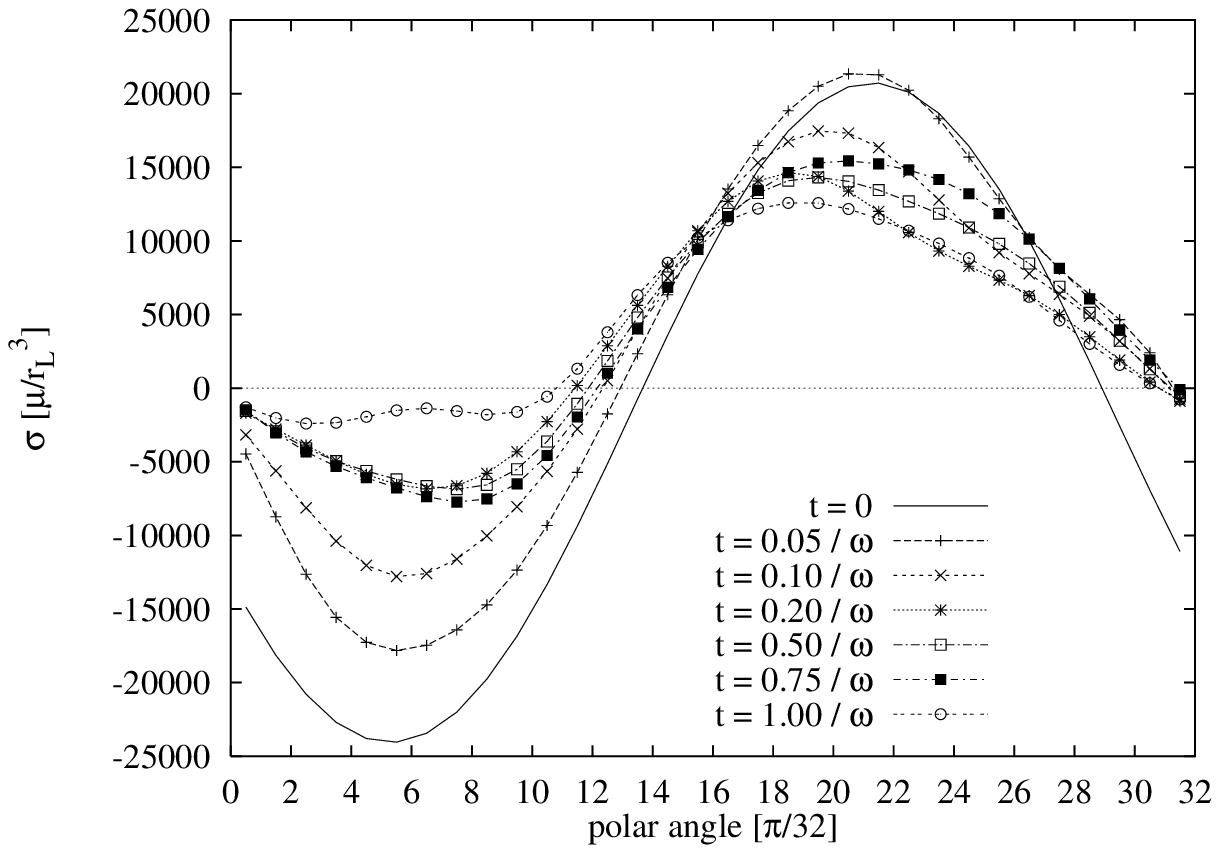}
\hfill
\includegraphics[width=7.2cm]{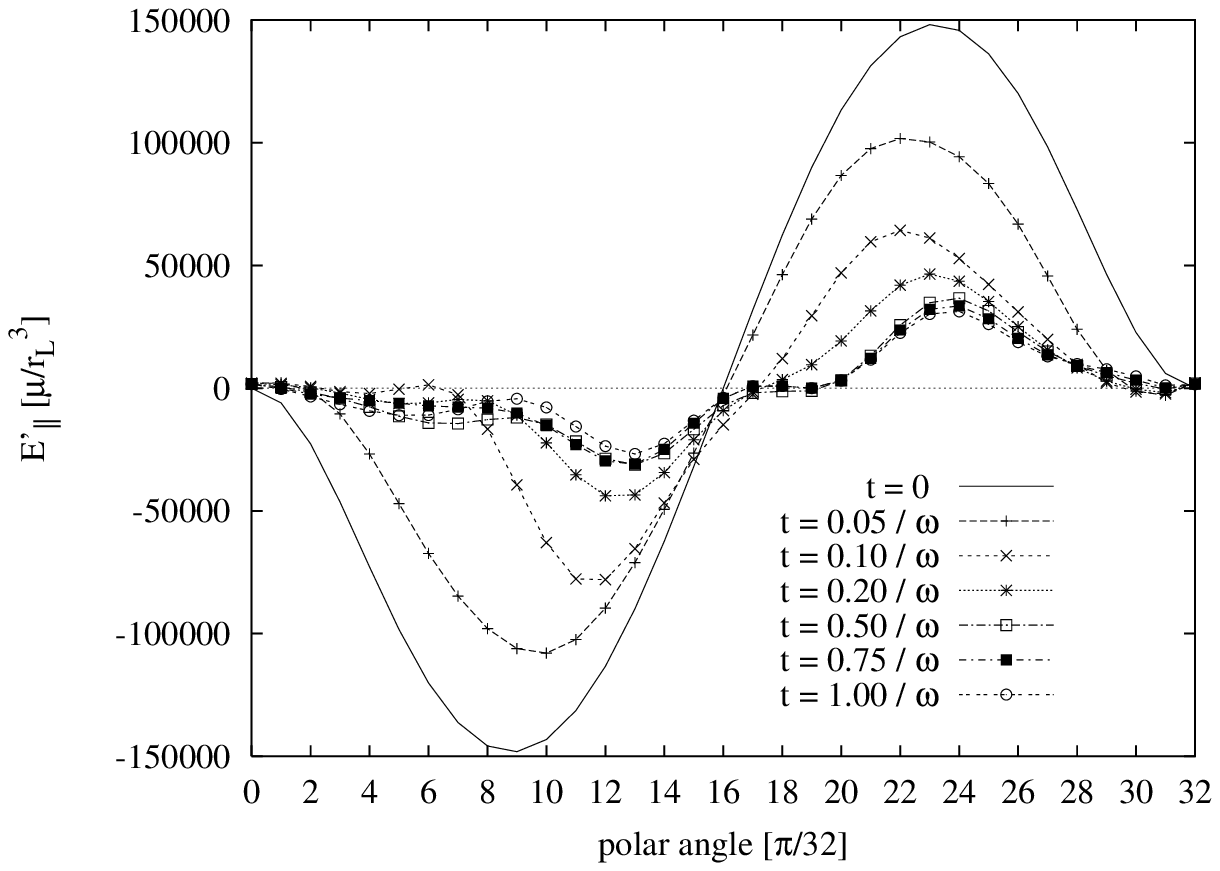}
\includegraphics[width=7.2cm]{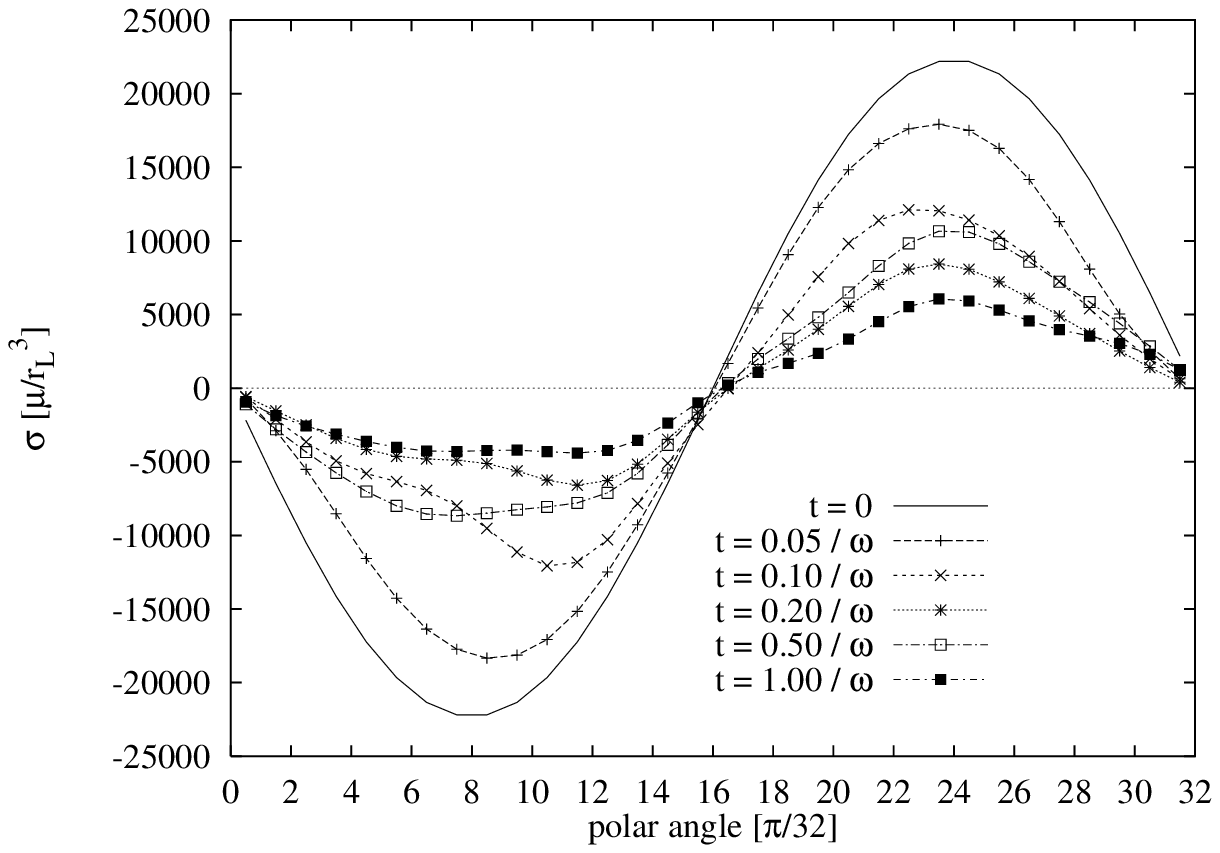}
\end{minipage}
\caption{Electric parallel field
 $E_{||}^{\pr} := {\rm sign}(\cos \psi) ({\bf E,B})/B$ on the surface of the 
sphere (left  plots) and the surface charge density (right plots) depending on
 the polar--angle relative to the rotation axis ($\l = 0$). From the top
 toword the bottom:$\chi=0^{\circ}$, $\chi=60^{\circ}$, $\chi= 90^{\circ}$.}
\label{fig_epar}     
\end{figure}

Likewise, diagramms on the left of figure \ref{fig_epar} show the development of the electric vector component parallel to the tangent to the respective magnetic field line over time on the stellar surface. Different curves, plotted against the polar angle, correspond to different values of the time coordinate. The three diagramms from the top to the bottom of figure \ref{fig_epar} are for different values of the inclination angle ranging from $\chi = 0^{\circ}$, $\chi = 60^{\circ}$ to  $\chi = 90^{\circ}$. In all these cases the electric vector component parallel to the tangent to the respective magnetic field line vanishes for all inclination angles within a time interval of about  $t \approx 0.2 \o^{-1}$, which indicates the formation of quasi stable magnetospheres. Analogous conclusions can be drawn from diagramms on the right of figure \ref{fig_epar} for the electric charge density on the surface of the sphere. 

\begin{figure}[h]
\centering
\begin{minipage}{14.4cm}
\includegraphics[width=7.2cm]{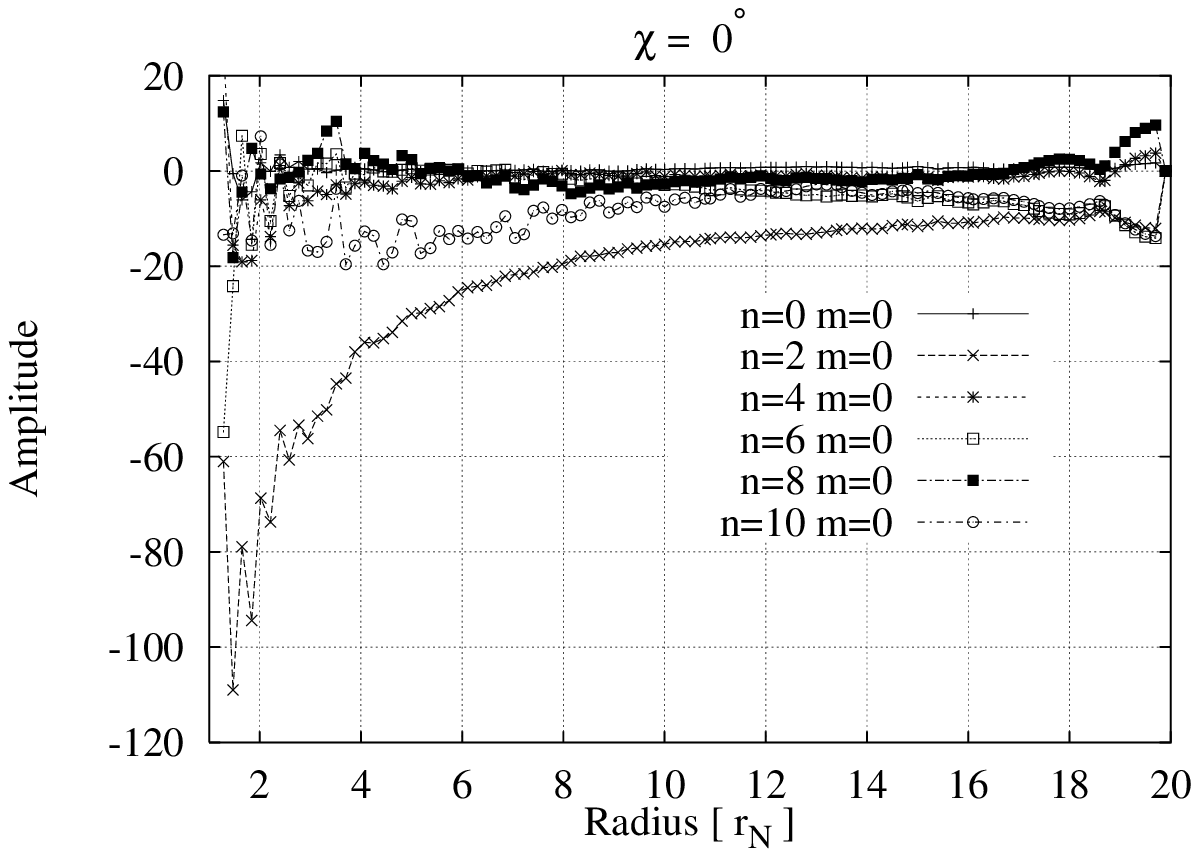}
\includegraphics[width=7.2cm]{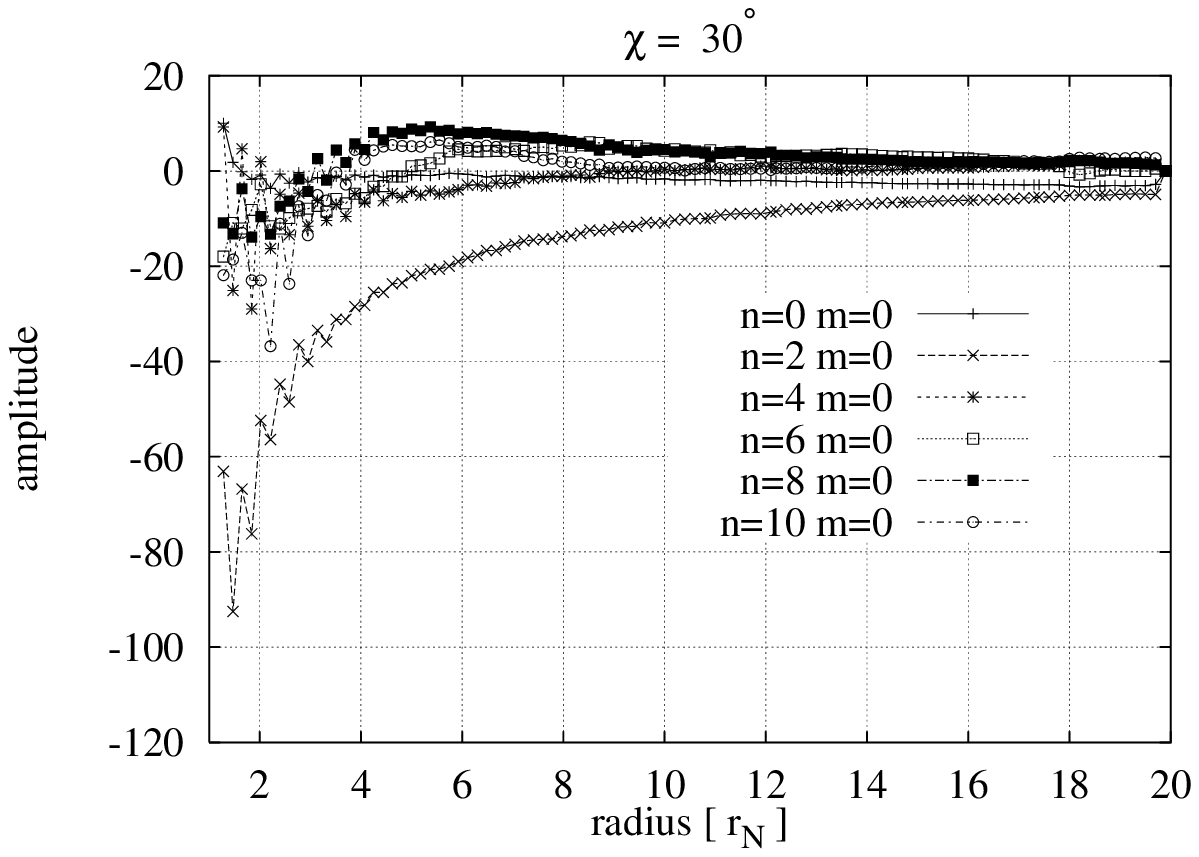}
\hfill
\includegraphics[width=7.2cm]{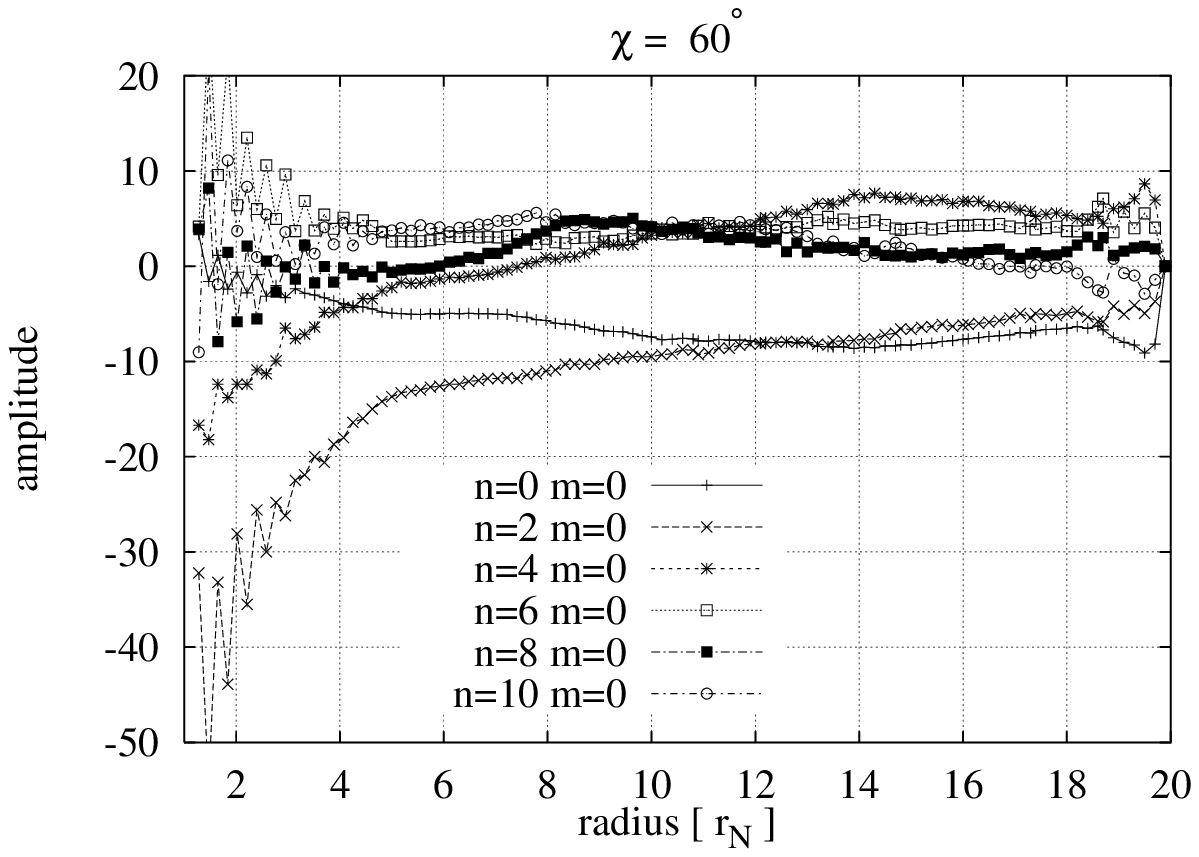}
\includegraphics[width=7.2cm]{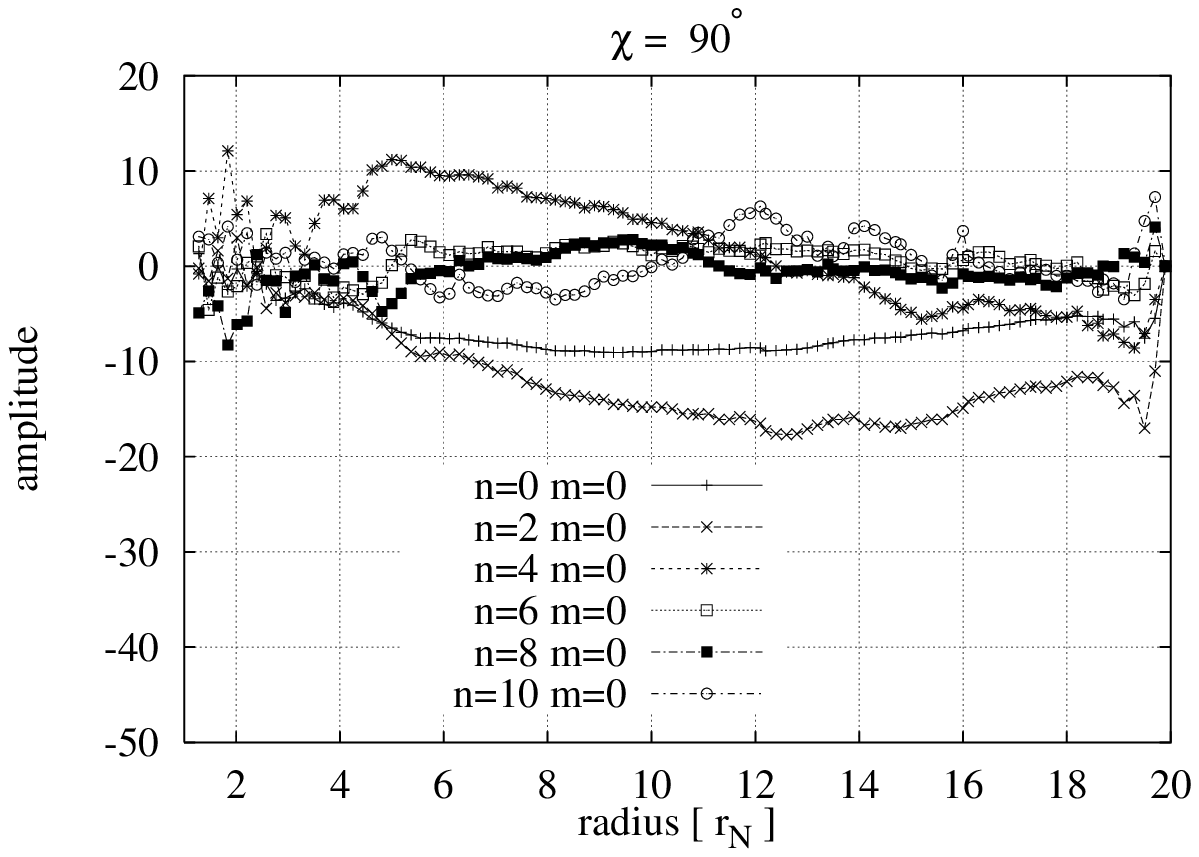}
\hfill
\includegraphics[width=7.2cm]{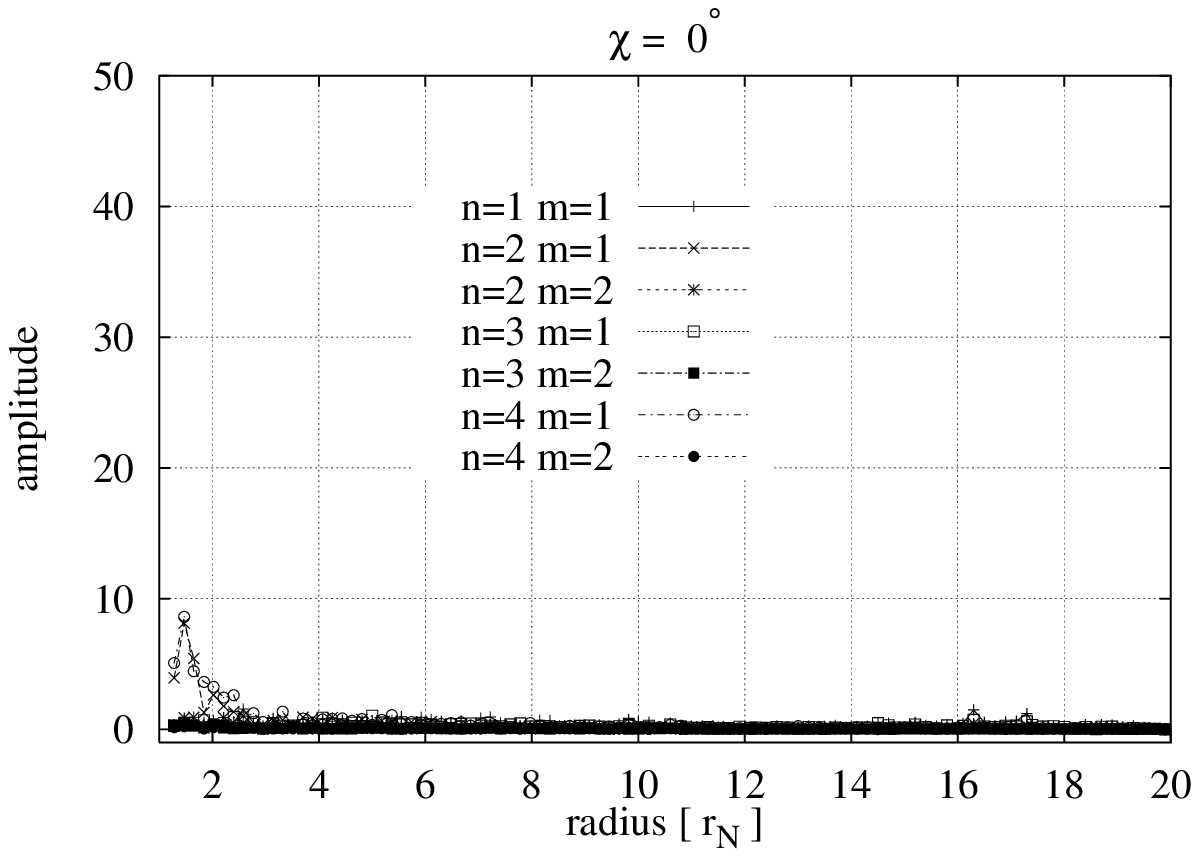}
\includegraphics[width=7.2cm]{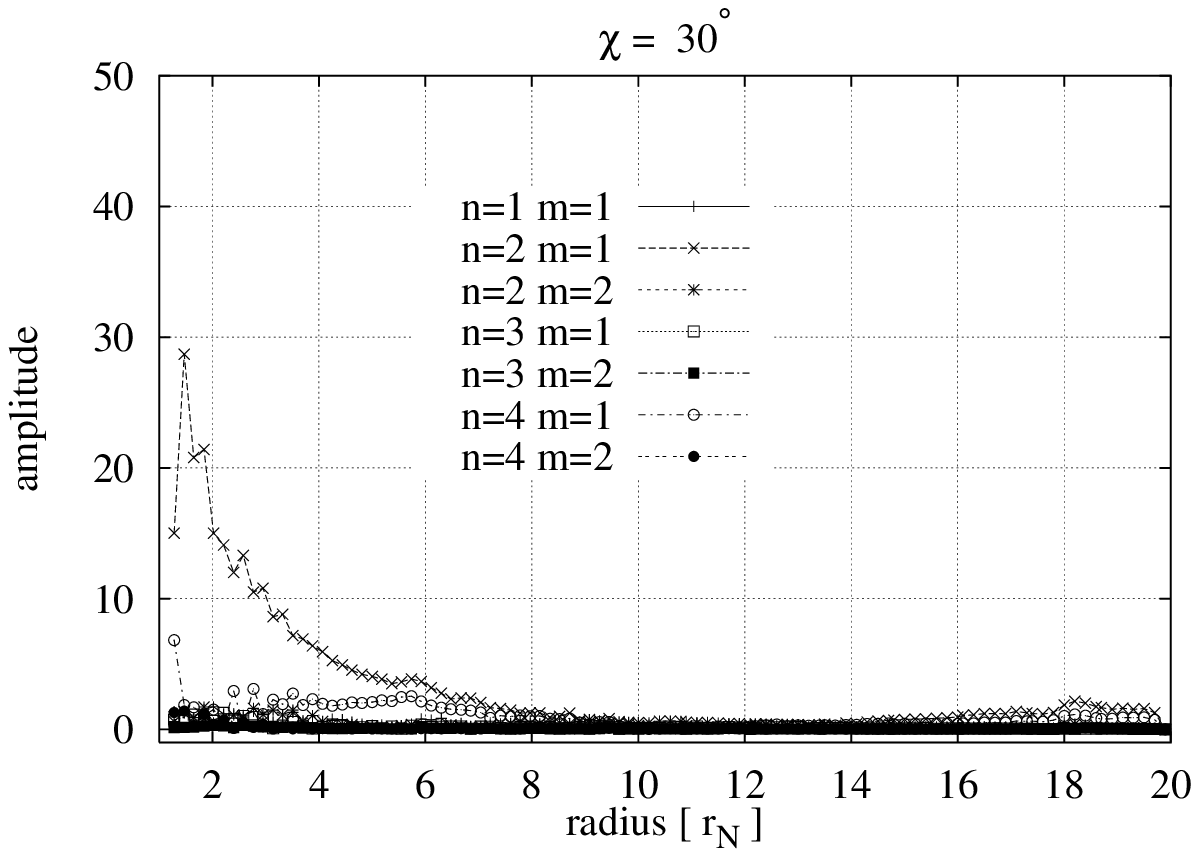}
\hfill
\includegraphics[width=7.2cm]{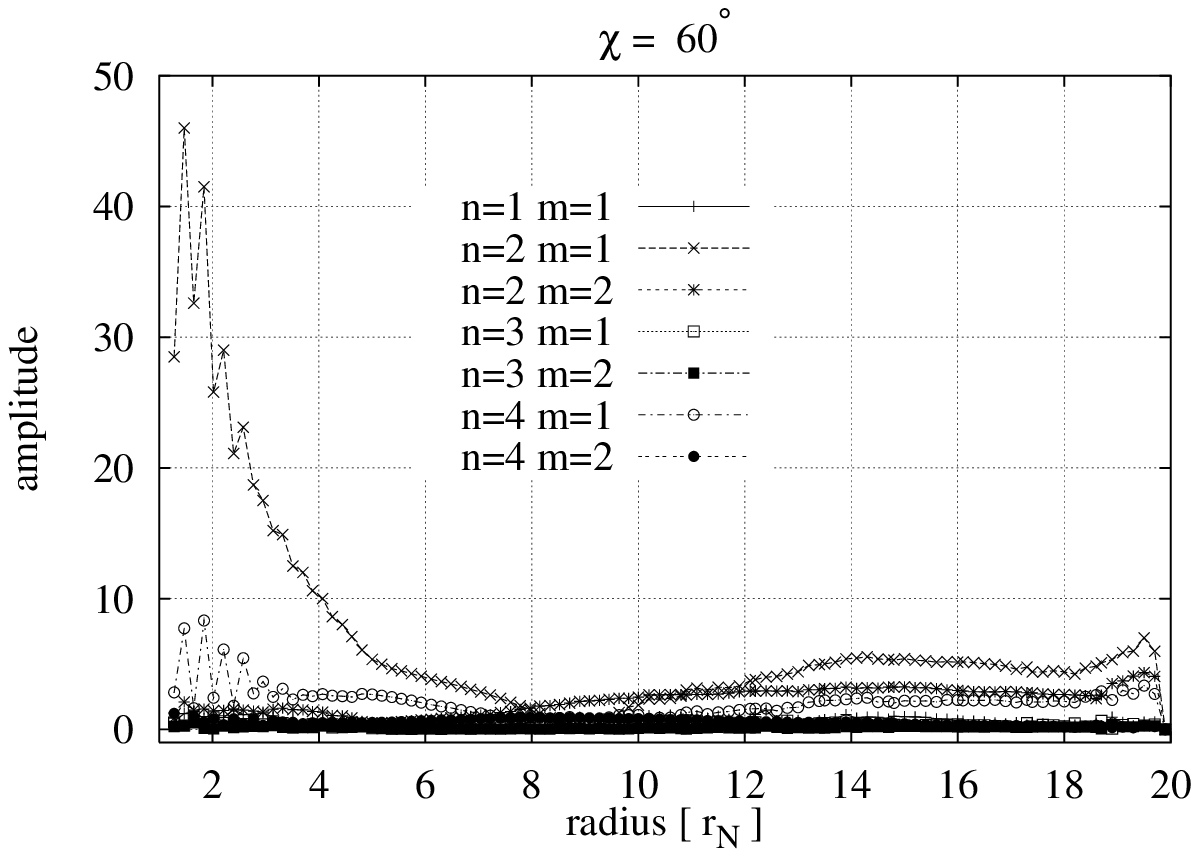}
\includegraphics[width=7.2cm]{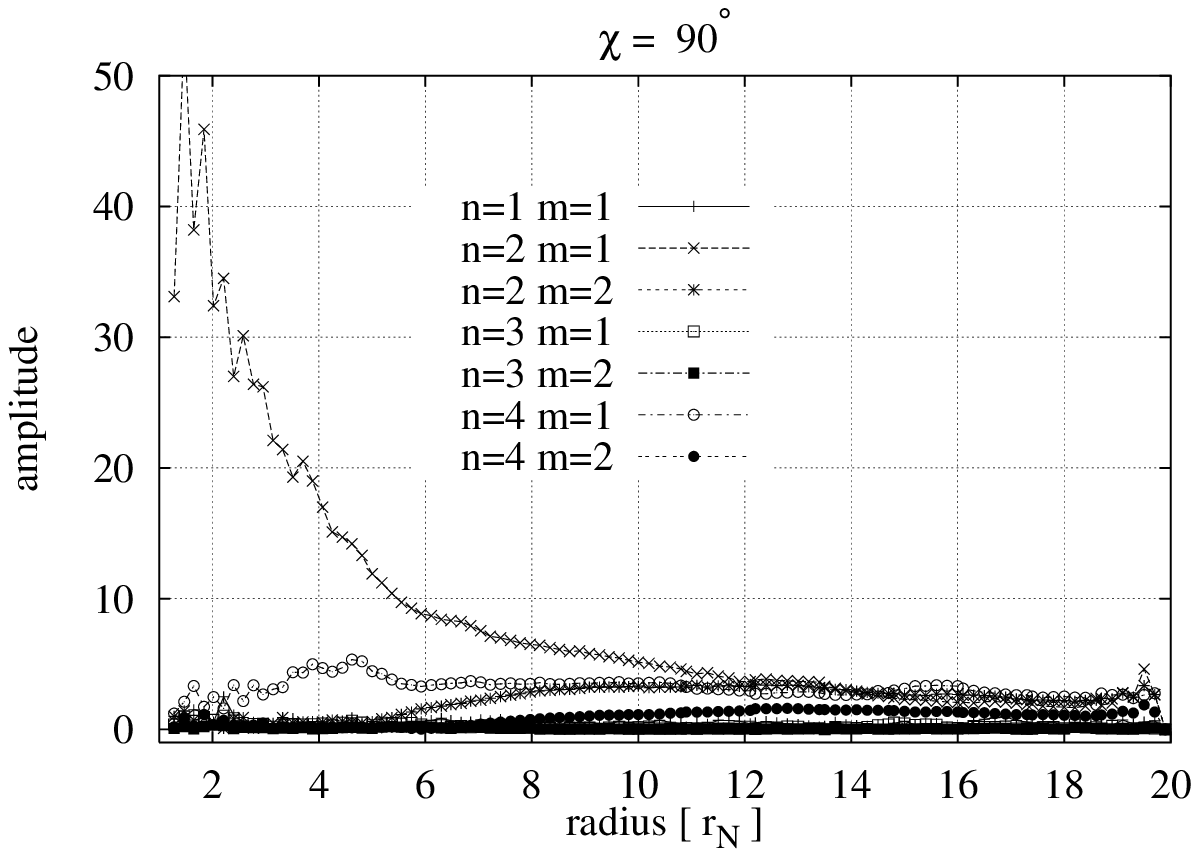}
\end{minipage}
\caption{Coeffients of the space charge density in the
 quasi--stationary case for different inclination angles. Upper rows: $m=0$; lower rows: $m\ne0$.}
\label{fig_moden}     
\end{figure}

For an analysis of spatial electric charge density inside the quasi-stationary magnetospheric particle distributions the corresponding even modes (normalized to $r^2$) in terms of coefficients of charge density (see chapter \ref{em}) as a function of the radial coordinate are shown in figure \ref{fig_moden}. Different curves are for different modes, while different diagramms indicate different inclination angles. In the special case of the aligned rotator the space charge density is described very well only by the quadropol mode $n=2, m=0$ in agreement with the predictions of the  Goldreich \& Julian--model as well as of the more recent work by \cite{neukirch}
\footnote{In the Goldreich \& Julian--model \cite{goldreich} a force--free magnetosphere ($({\bf E,B})=0$) and  a co-rotating plasma inside the light--cylinder are assumed. With ${\bf E} = - [\Bb^{{\rm koro}},{\bf B}]$ and $\Bb^{{\rm koro}} = (r/r_L) \sin \t {\bf e}_{\p}$, the electric field outside the sphere results in: ${\bf E} = - \frac{\m k^3}{( k r)^2}( - \sin^2 \t \, {\bf e}_{r} +  2 \cos \t \sin \t \, {\bf e}_{\t})$. This field is caused by the charge density $\r_{{\rm GJ}} = - \frac{\m}{\pi r_L} \frac{1}{r^3} P_2(\cos \t)$ and an additional sphere--monopol by $q_s = \frac{2}{3} \frac{\m}{r_L}$. The electric potential is given by $\label{A0_GJ} A_0 = A_0^{{\rm mono}} + A_0^{{\rm quadru}} = \frac{2}{3} \frac{\m}{r_L} \frac{1}{r}(\,1 - P_2(\cos \t)\,)$.}. For increasing inclination angle $\chi$ other modes gain more and more importance, especially those with $m\ne 0$ which are responsible for non--axially symmetric constibutions, as shown in figure \ref{fig_moden}. 

\begin{figure}[h]
\centering
\begin{minipage}{12.4cm}
\vspace{-1.2cm}
\includegraphics[width=6cm]{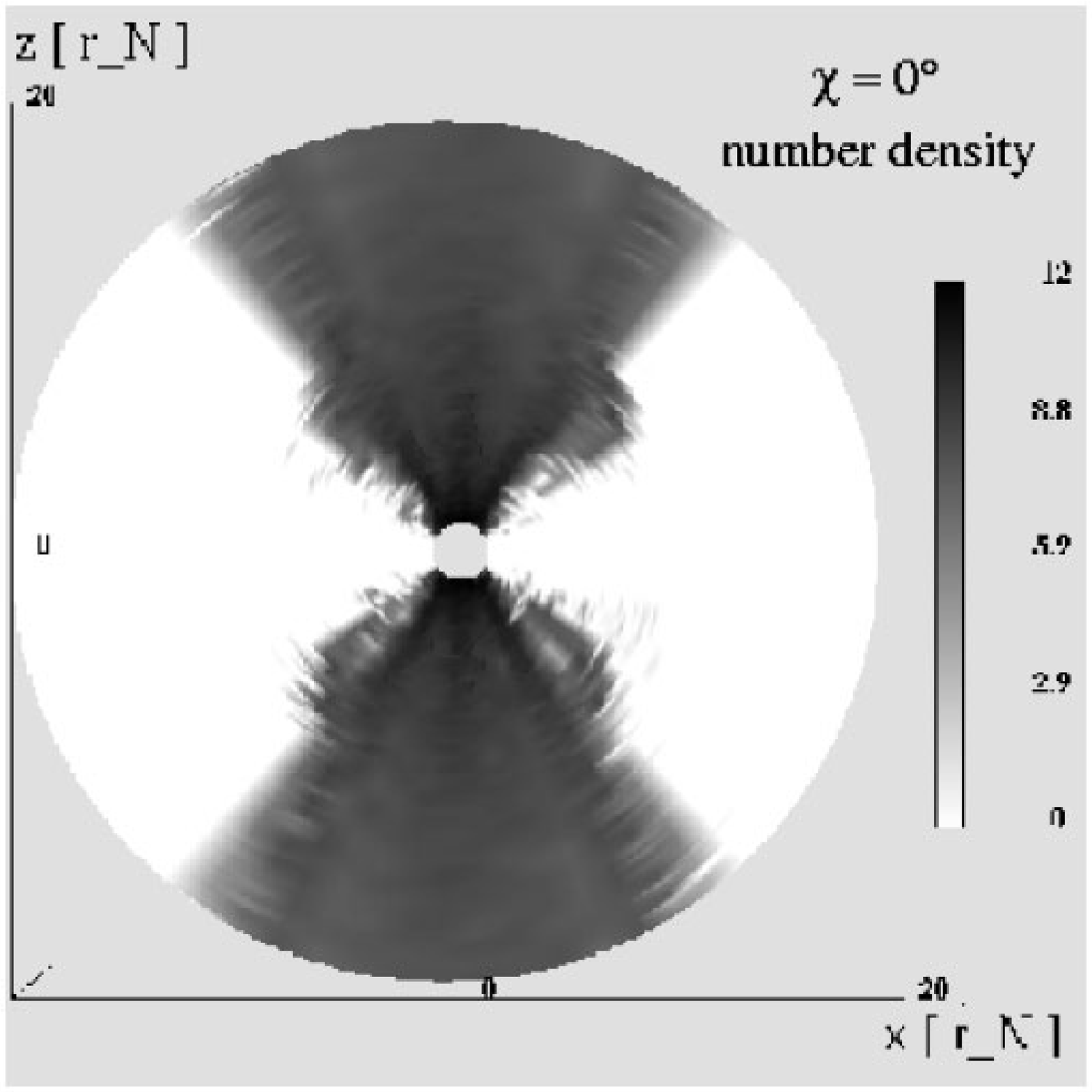}
\includegraphics[width=6cm]{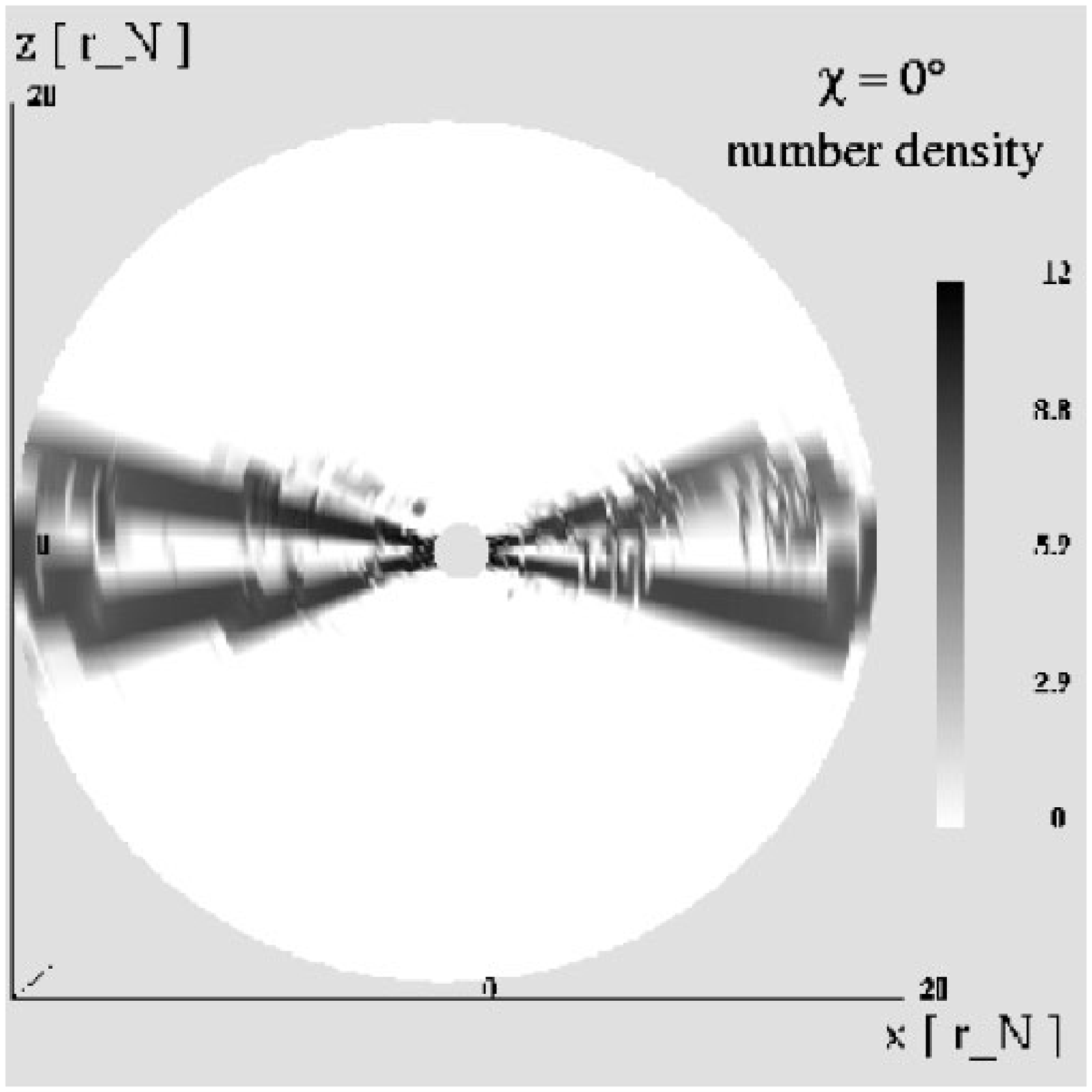}
\hfill
\vspace{-1.2cm}
\includegraphics[width=6cm]{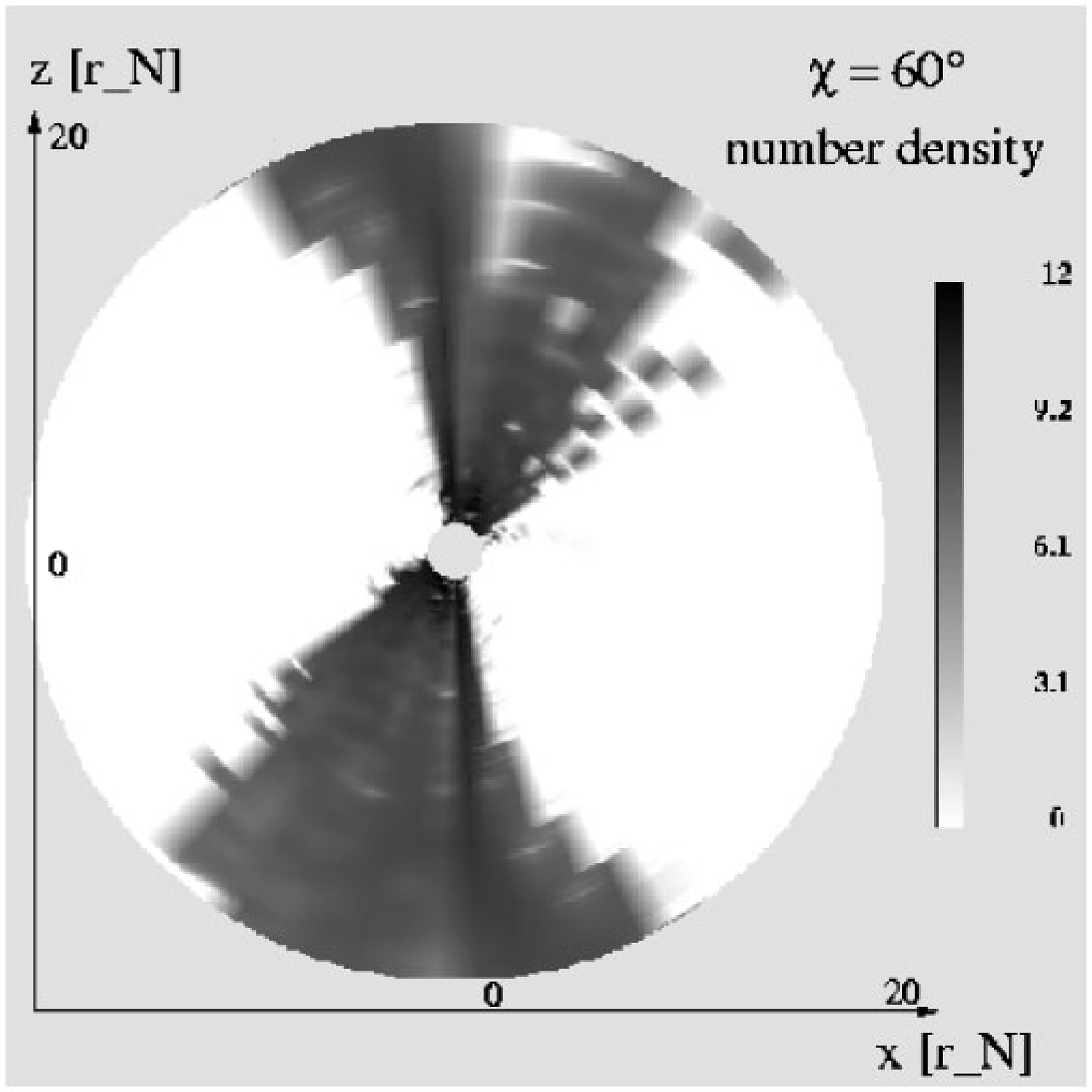}
\includegraphics[width=6cm]{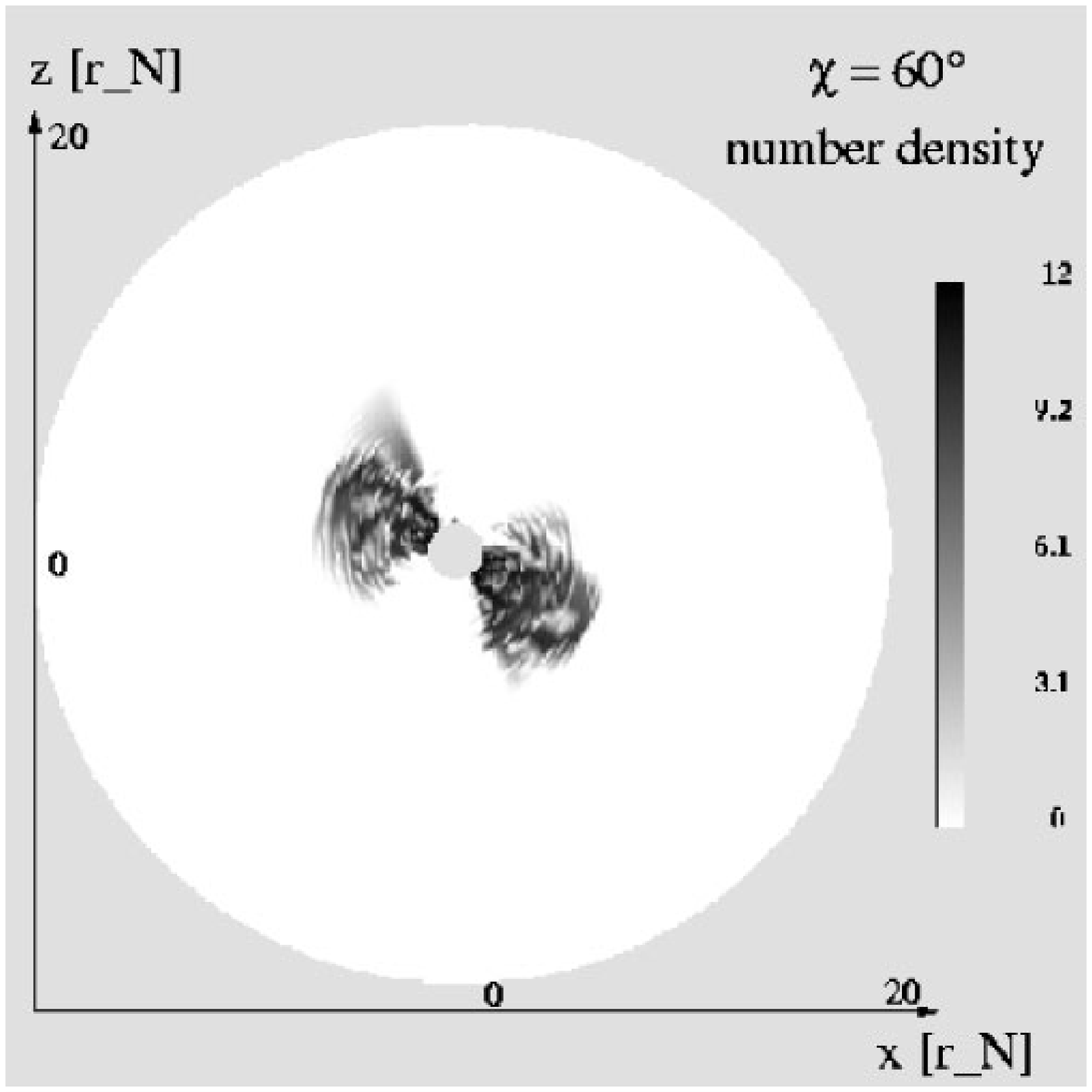}
\hfill
\vspace{-1.2cm}
\includegraphics[width=6cm]{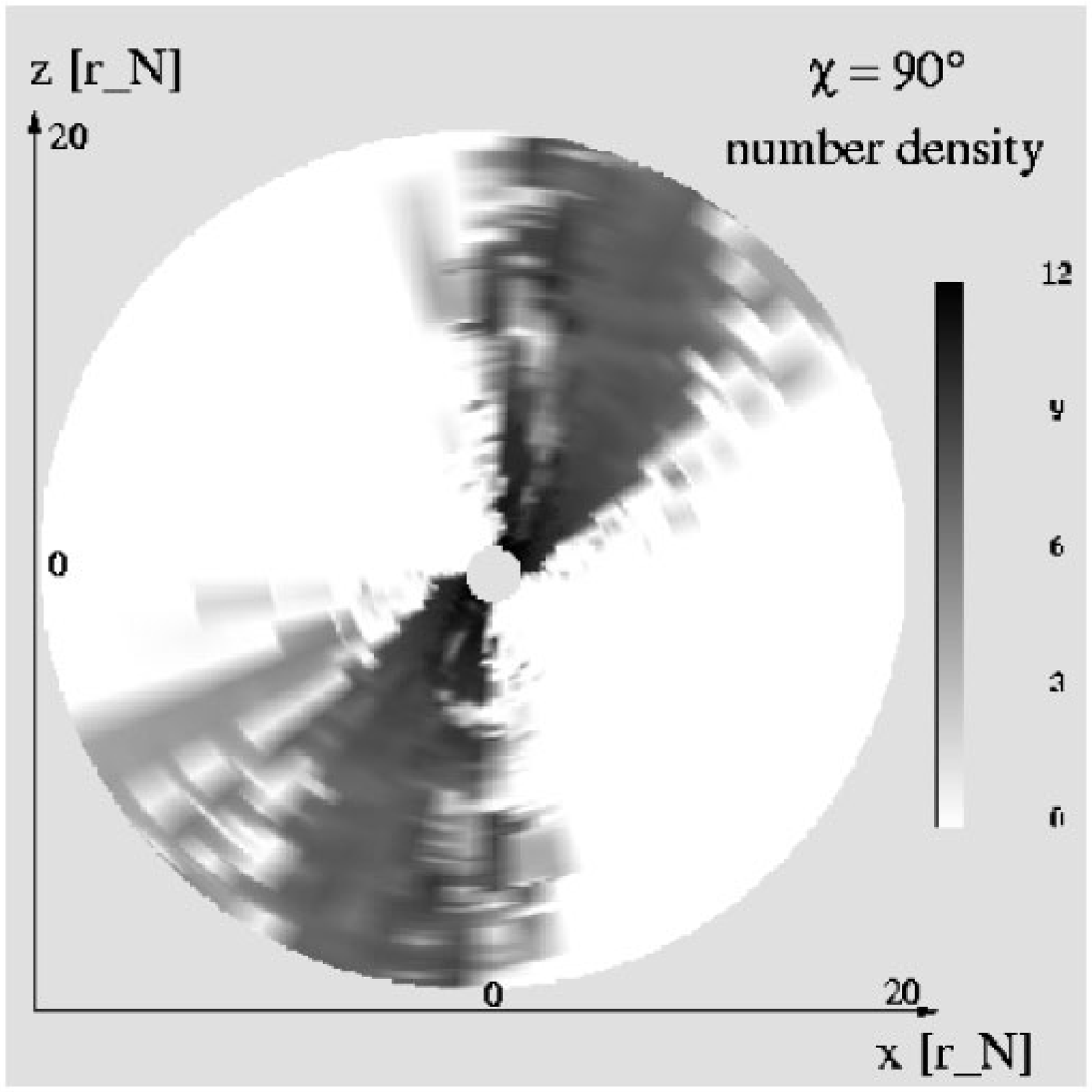}
\includegraphics[width=6cm]{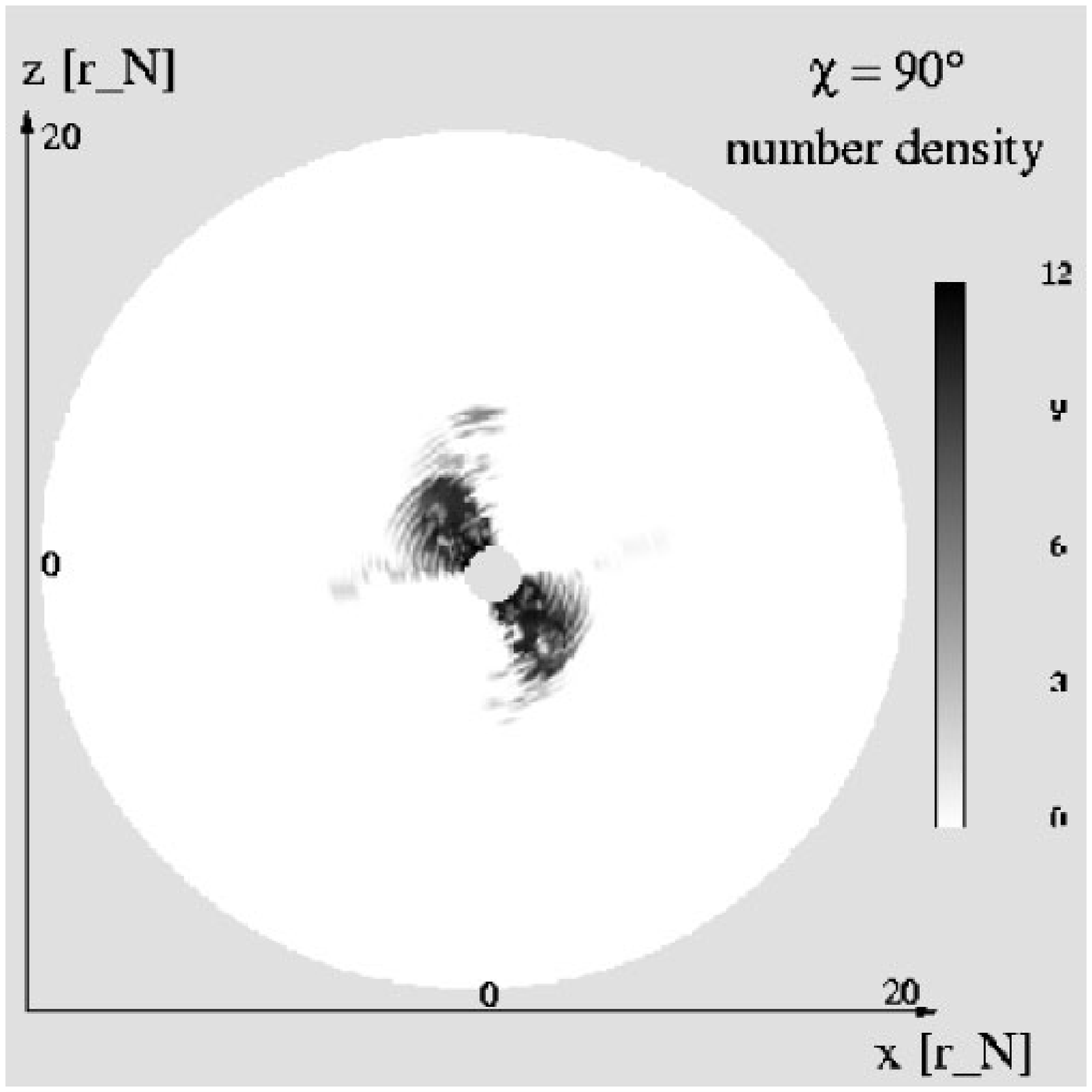}
\end{minipage}
\vspace{-1.2cm}
\caption{Particle density [$cm^{-3}$] of the electron fluid (left fig.~)
 and  proton fluid (right fig.~) mapped over $\log_{10}( \r\,cm^3 + 1)$ after a
 rotation $\simeq 57.3^{\circ}$ for different inclination angles.}
\label{fig_rho1}       
\end{figure}

For a better vizualization of spatial electric charge distribution inside the quasi-stationary configuration, in figure \ref{fig_rho1} the number densities of the electron fluid (on the left) and of the proton fluid (on the right) within the plane spanned by the magnetic and the
 rotation axis are shown for different inclination angles. 

The structures of the resulting quasi-stationary magnetospheres are dominated by the force--free surfaces for all inclination angle and are completly charge seperated devided by regions of vanishing particle number density, often refer as 'vacuum gaps'. 

In the case of the inclined and orthogonal rotator, $0 < \chi \le \frac{\pi}{2}$, electrons are collected between the rotation and the magnetic axis, nearby and in the plane spanned by these axes. The fluid includes the rotation axis, except for $|\chi - \frac{\pi}{2}| \approx 0$. Protons are collected between the equator plane relative to the rotation axis and the equator plane relative to the magnetic axis, once again nearby and in the plane spanned by these axes. Given $\frac{\pi}{2} < \chi \le \pi$ the sign of the particles changes.

We found corotation (relative to the surface of the sphere) for all inclination angle. Particle number densities inside these clouds, for the standard set of parameters, typically range up to $10^{12}\,{\rm cm}^{-3}$.

\begin{figure}[h]
\centering
\begin{minipage}{15cm}
\vspace{-1.5cm}
\includegraphics[width=5cm]{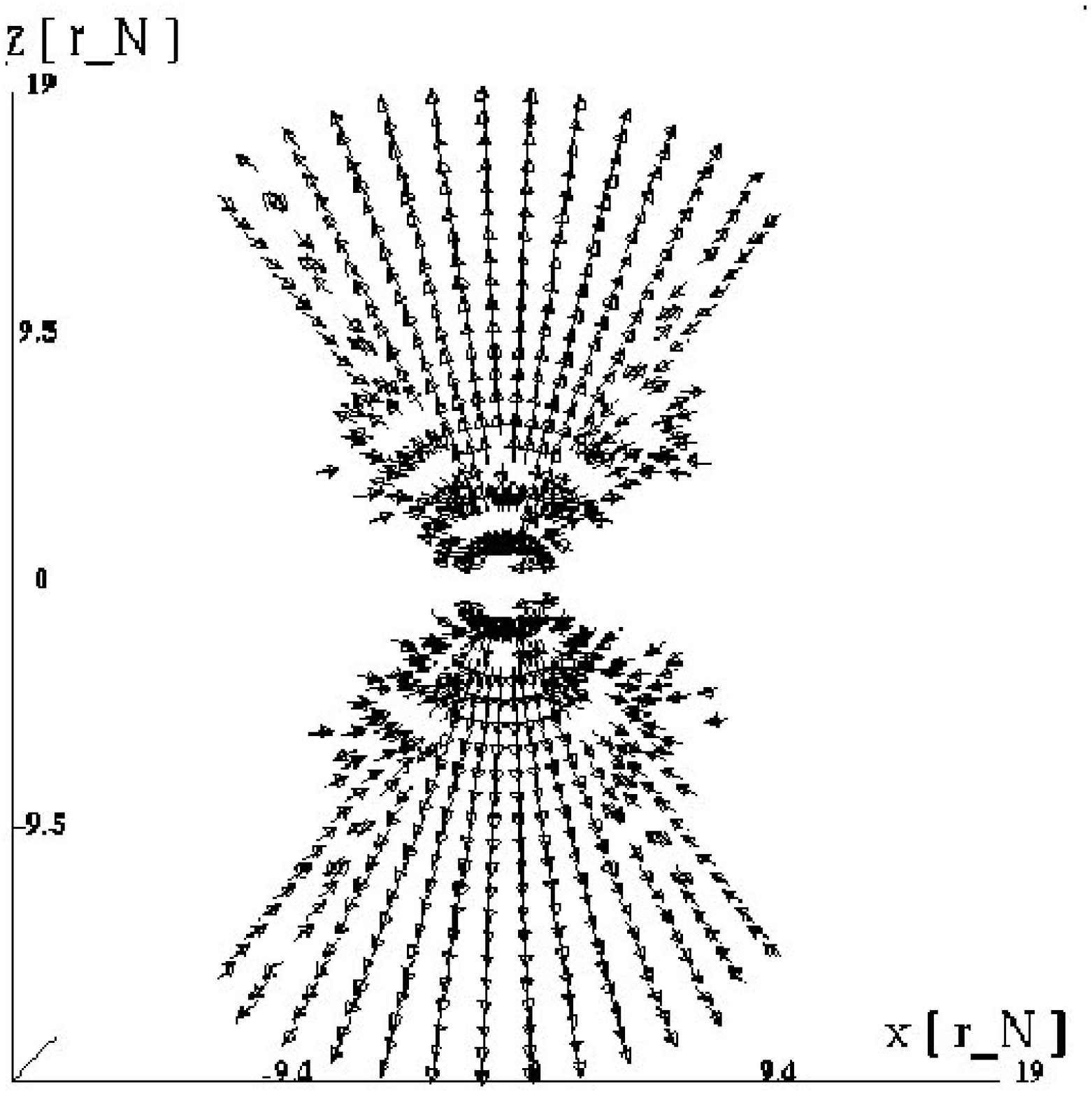}
\hspace{0.5cm}
\includegraphics[width=5cm]{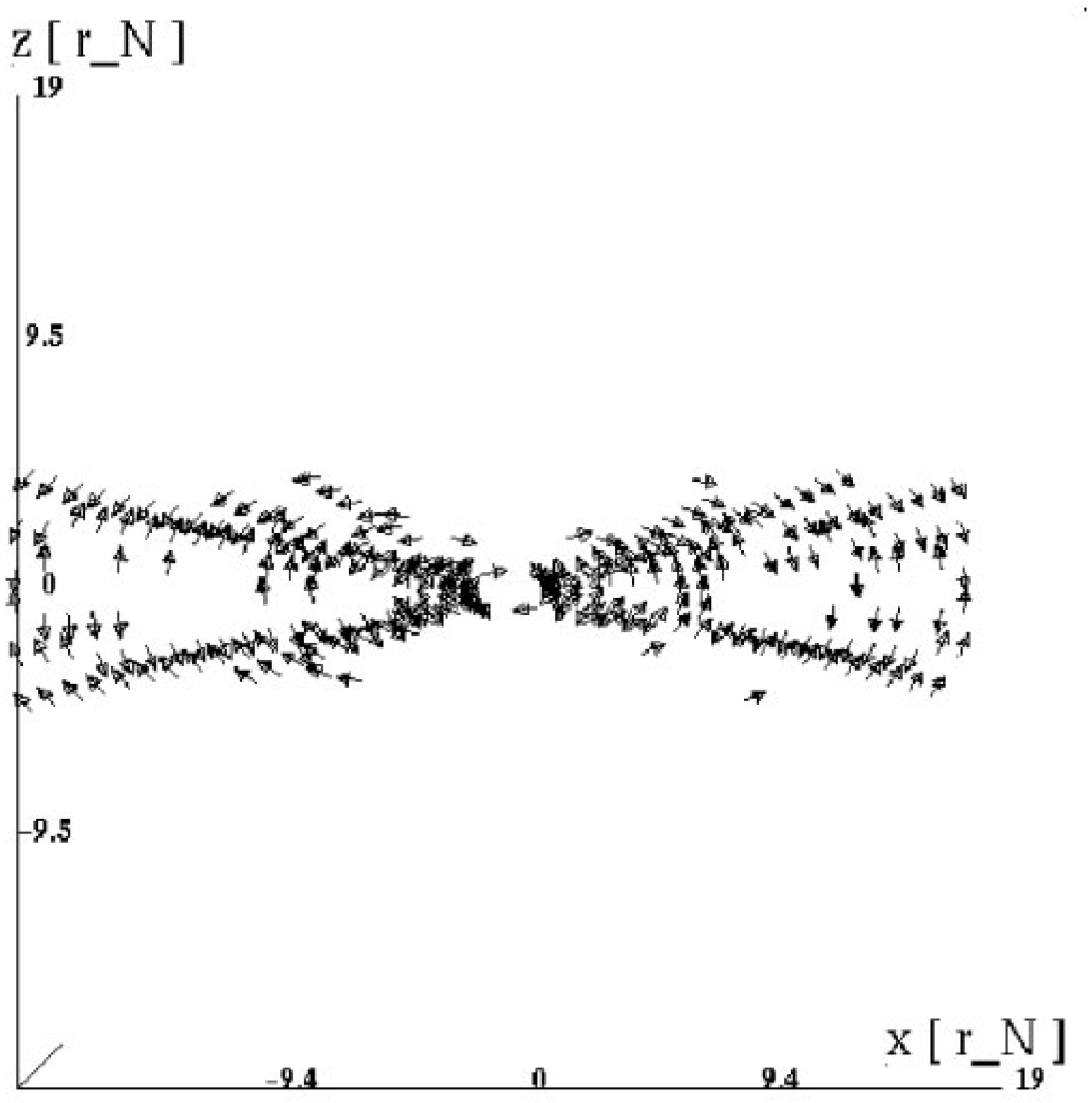}
\hspace{-1.5cm}
\includegraphics[width=5cm]{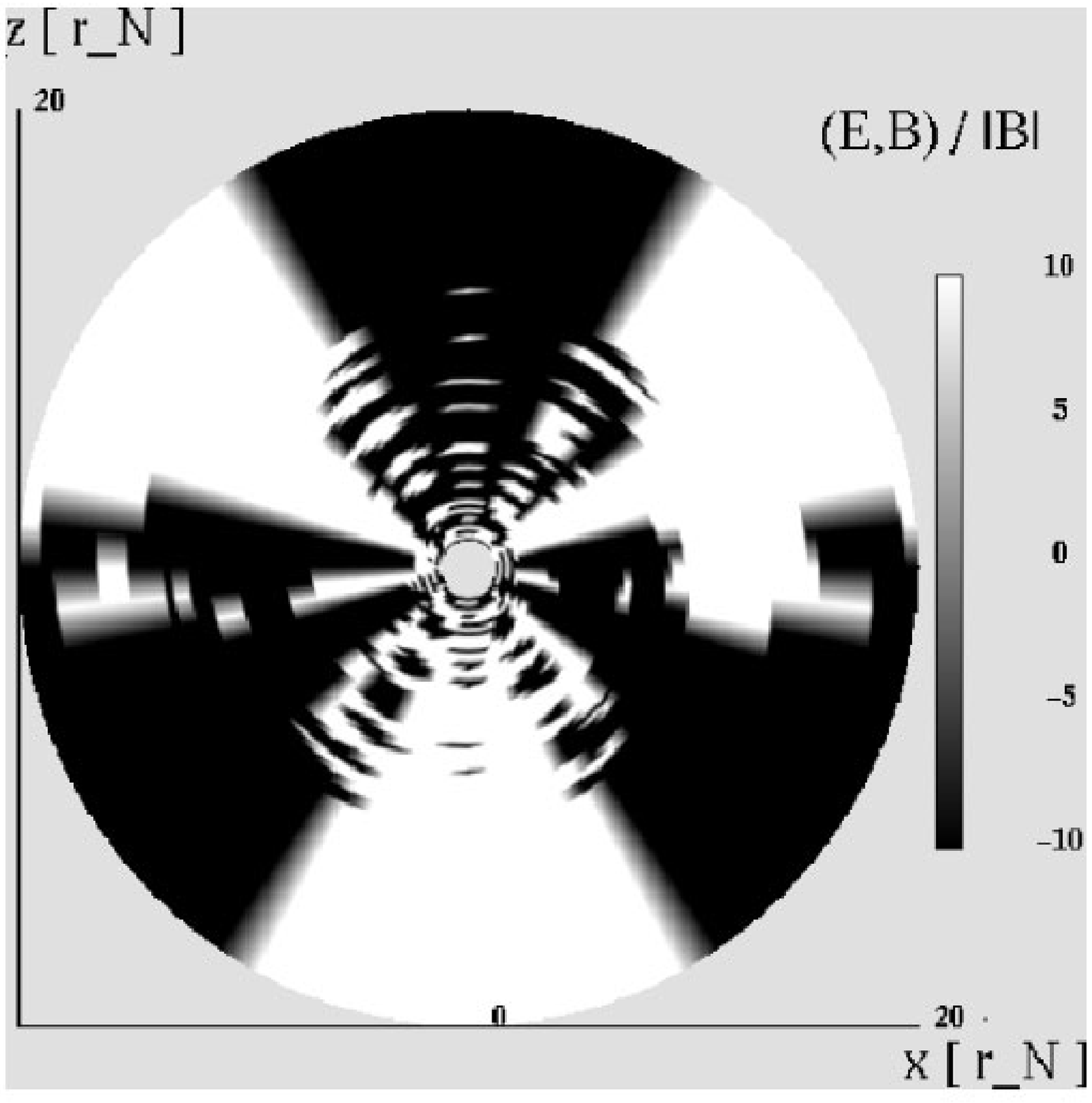}
\hfill
\vspace{-1cm}
\includegraphics[width=5cm]{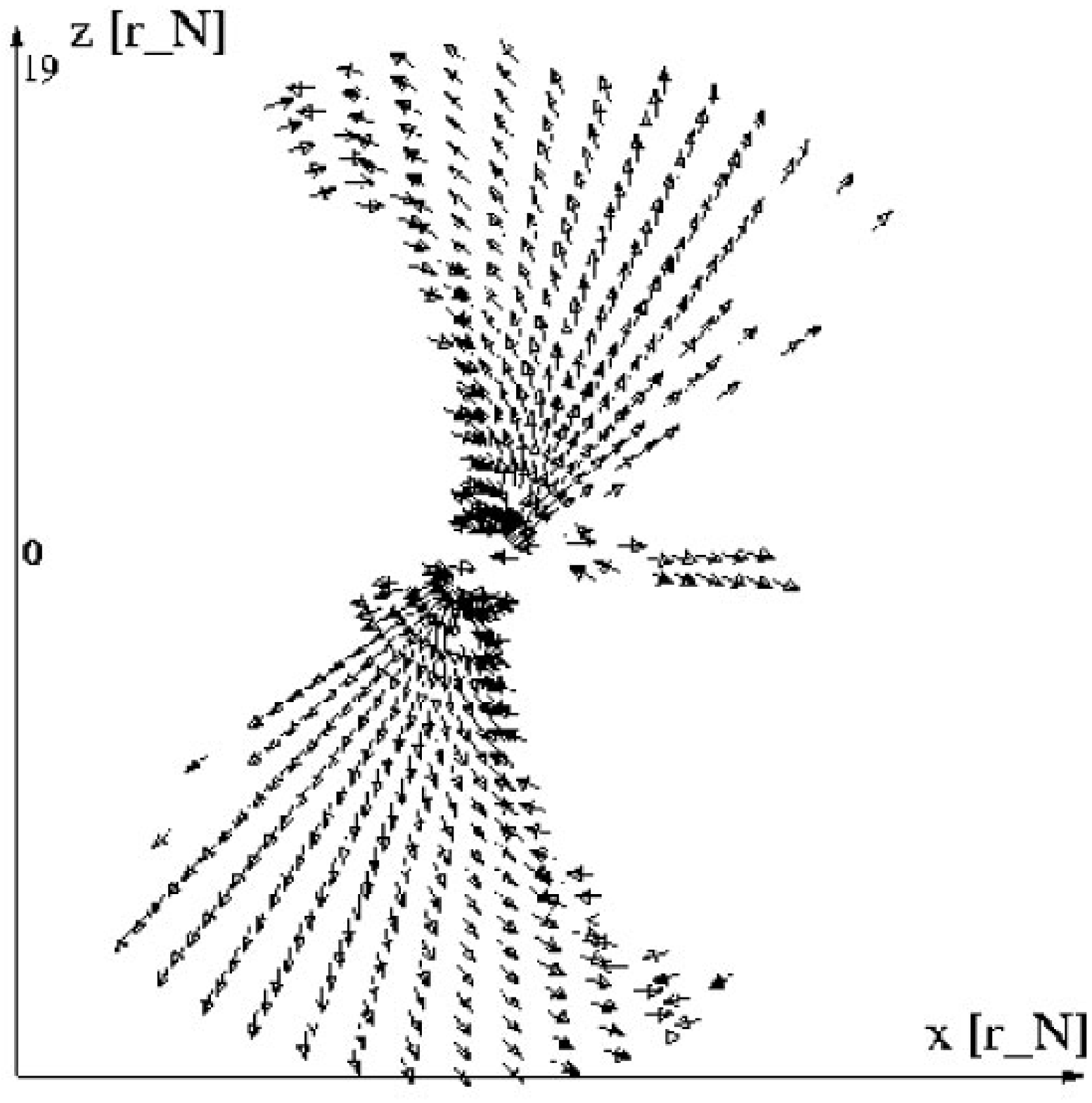}
\hspace{0.5cm}
\includegraphics[width=5cm]{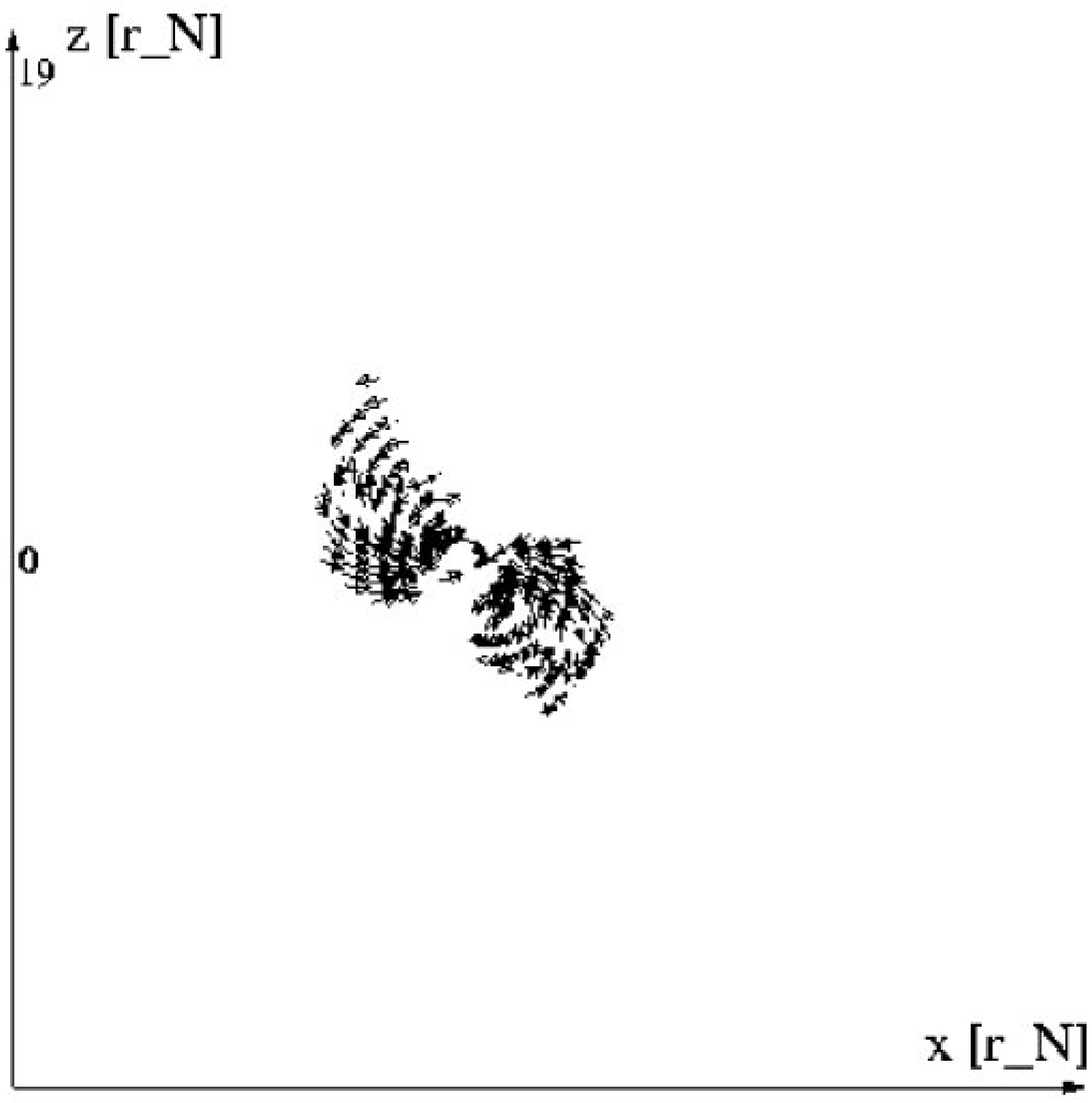}
\hspace{-1.5cm}
\includegraphics[width=5cm]{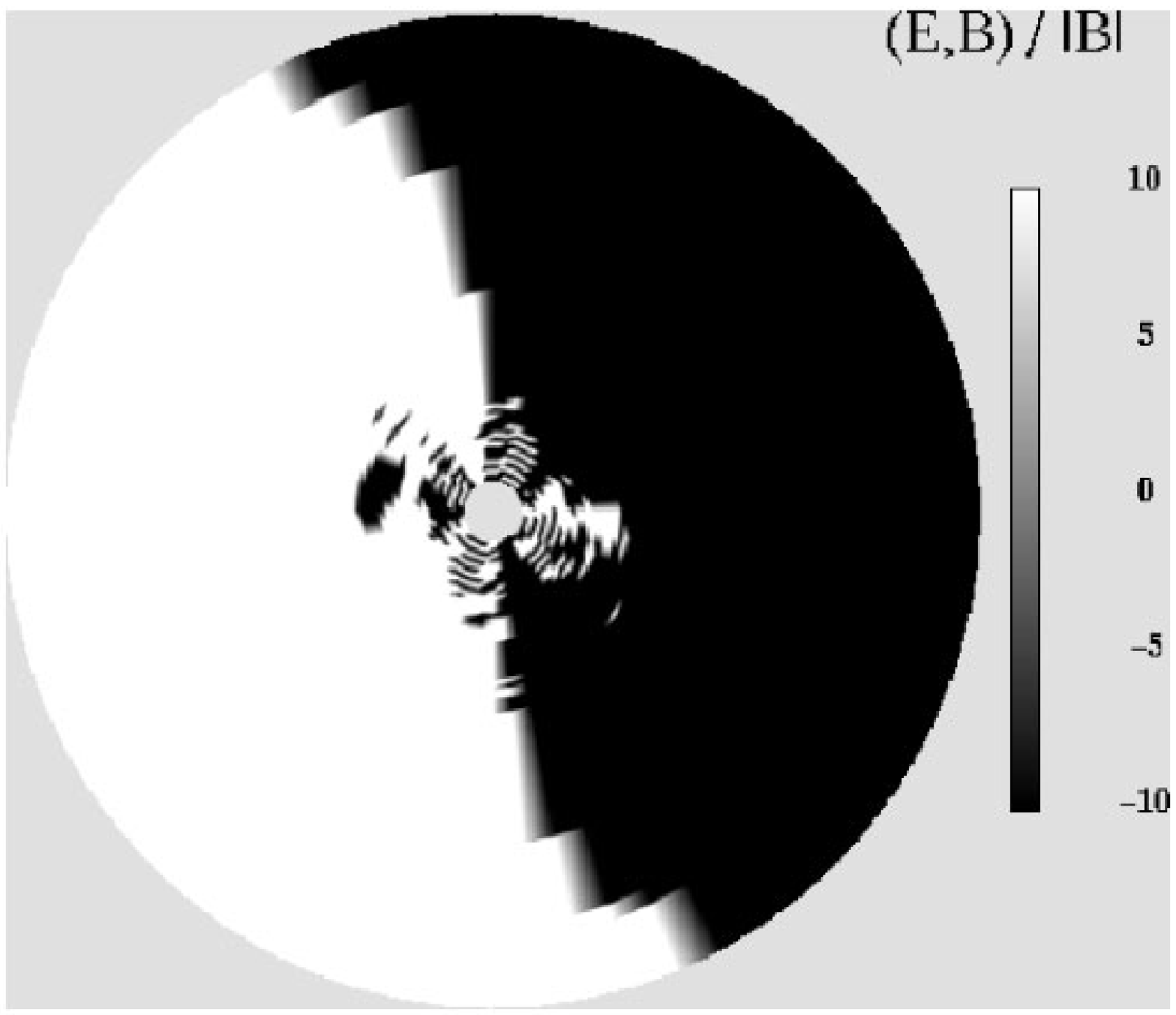}
\hfill
\vspace{-1cm}
\includegraphics[width=5cm]{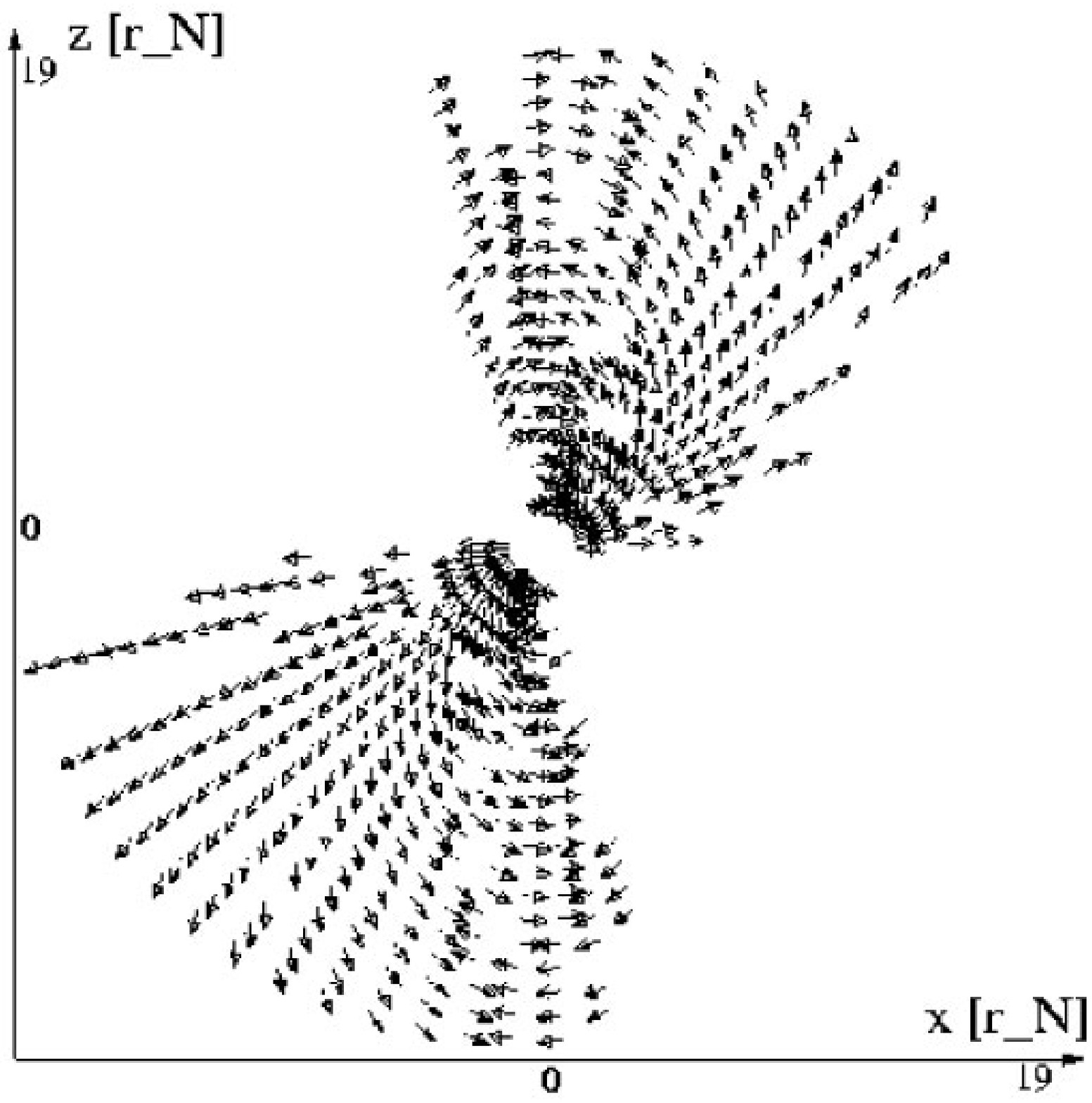}
\hspace{0.5cm}
\includegraphics[width=5cm]{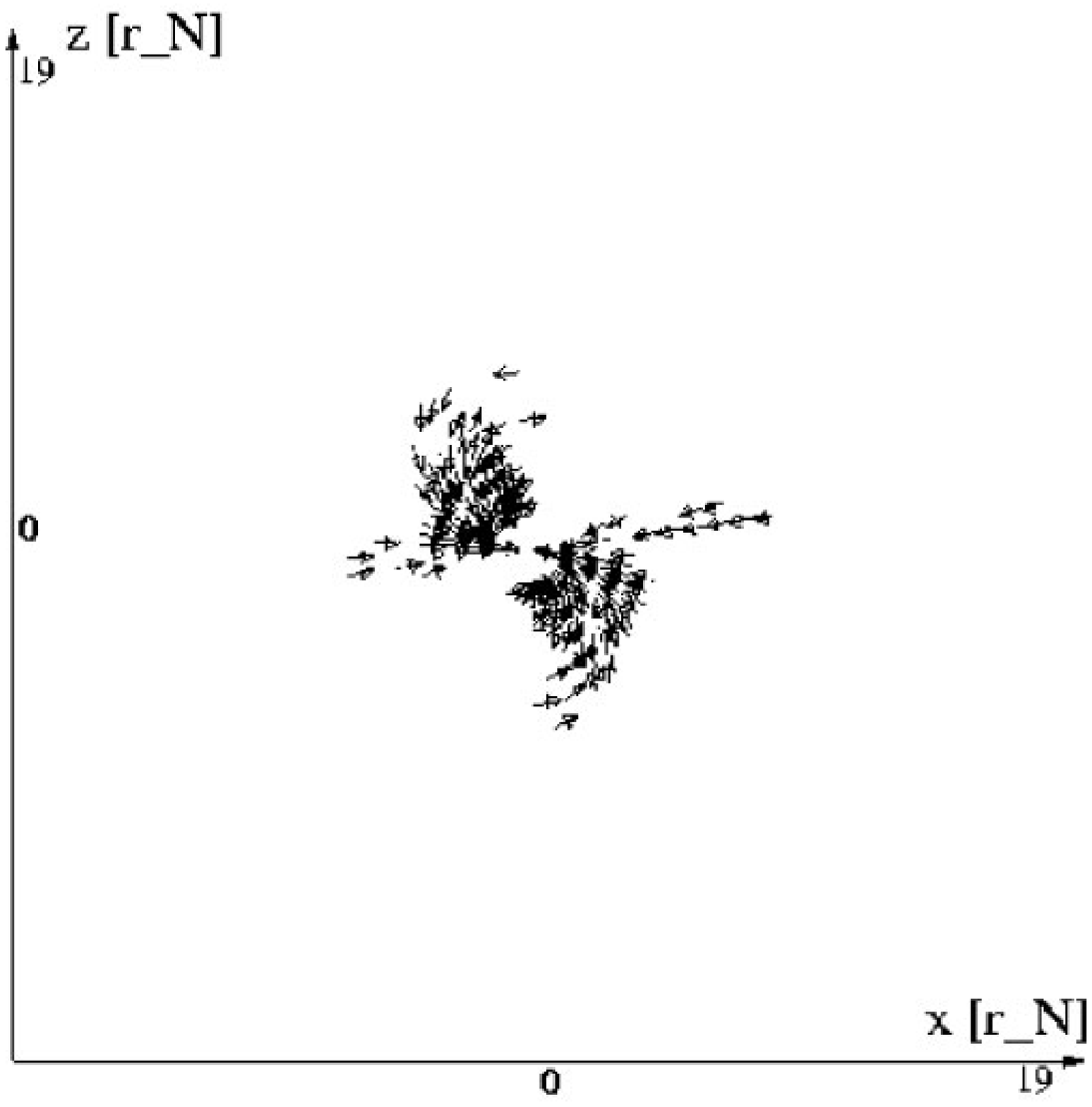}
\hspace{-1.5cm}
\includegraphics[width=5cm]{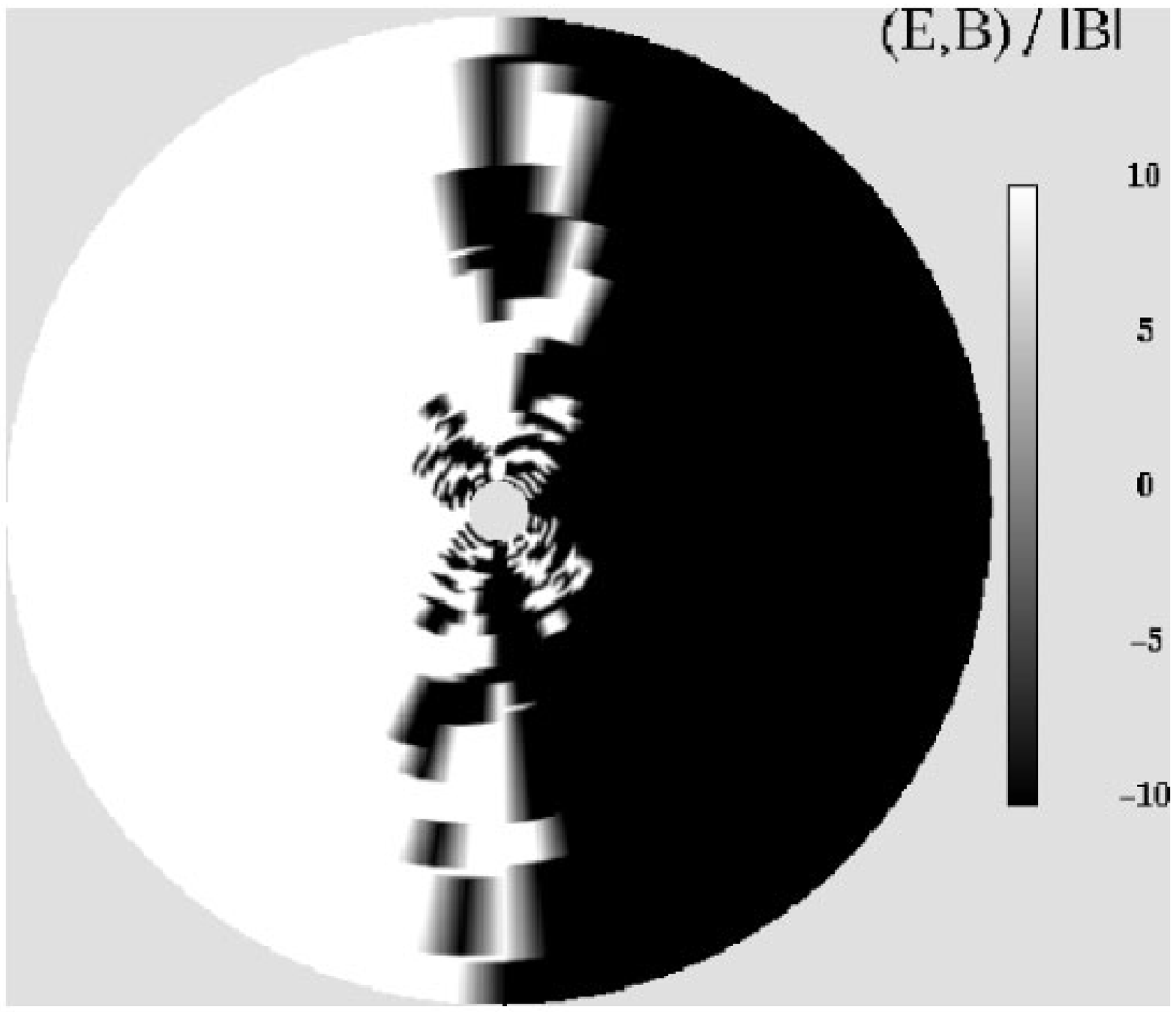}
\end{minipage}
\vspace{-1cm}
\caption{ Fig.~left: Normalized velocity field of the electron fluid. Fig.~middle: Normalized velocity field of the proton fluid. Fig.~right: Electric parallel field $E_{||}=({\bf E,B})/B$. $|E_{||}| > 10$ is mapped to 
$\pm 10$ in order to show the change in the sign. From top to bottom: $\chi=0^{\circ}$, $\chi=60^{\circ}$, $\chi=90^{\circ}$}
\label{fig_u_esb}   
\end{figure}

\begin{figure}[h]
\centering
\begin{minipage}{15cm}
\includegraphics[width=7.5cm]{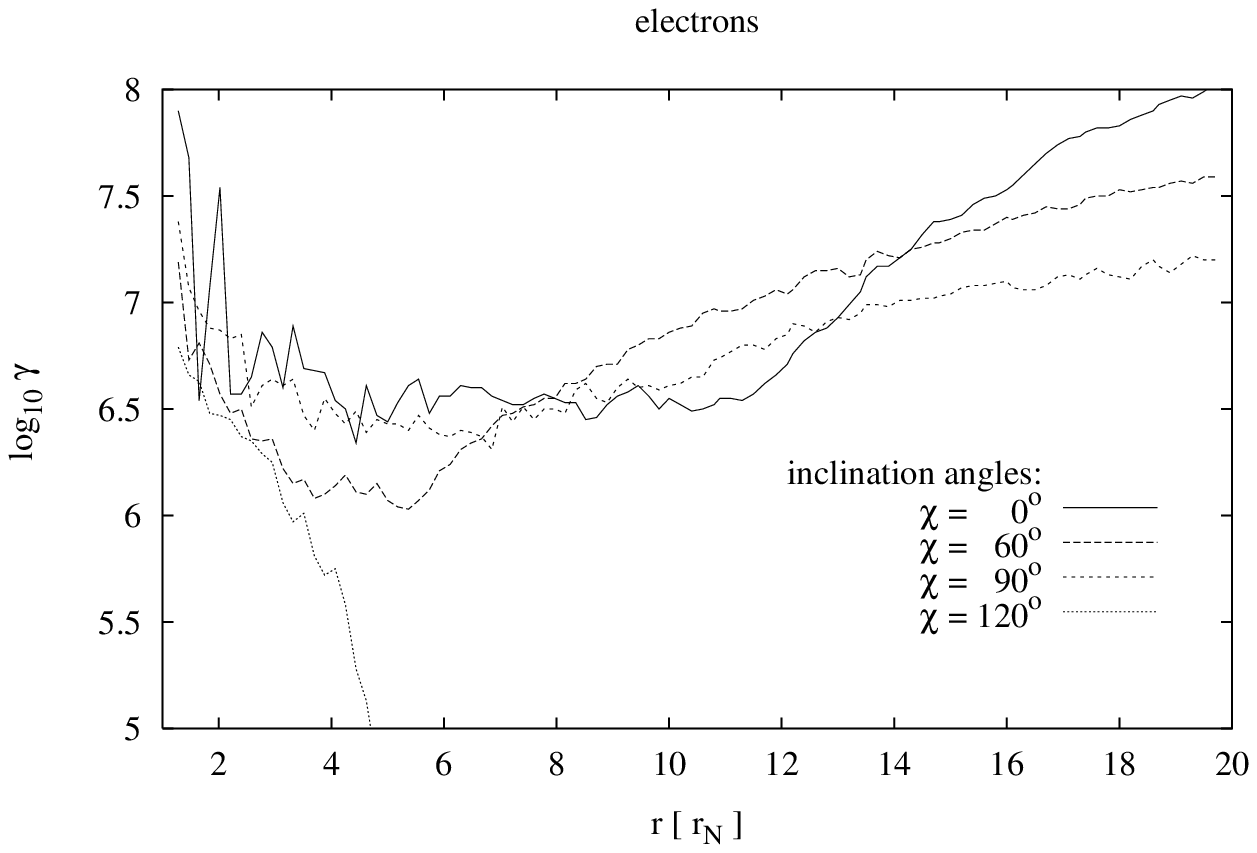}
\includegraphics[width=7.5cm]{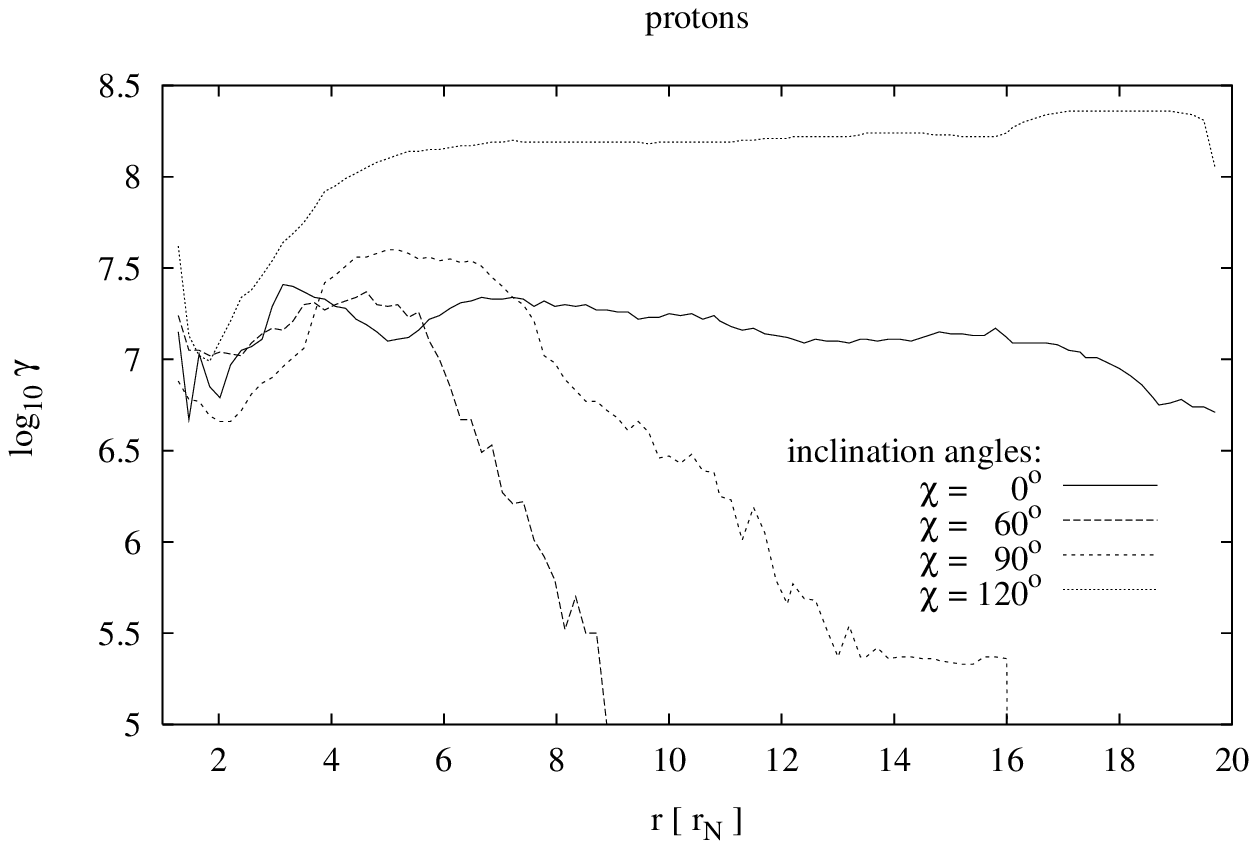}
\end{minipage}
\caption{Averaged  Lorentz--factors per radial sphere depending on
 the radius for different inclination angles.}
\label{fig_gamma}     
\end{figure}

For a discussion on currents and  particle acceleration within these clouds normalized velocity fields of electrons and protons are considered in figure \ref{fig_u_esb} for various inclination angles. Furthermore the projection of the electric vector onto the tangent to the magnetic field lines, 
$({\bf E,B})/ |{\bf B}|$, is shown. The direction of the velocity field correlates monotonously with the sign of these projection of the electric vector. The averaged  Lorentz--factors for electrons  and protons are shown in figure \ref{fig_gamma}. Average values are calculated for spherical shells (the radius of a given shell is mapped on the abscissa). Different curves indicates different values of the inclination angle.

In the case of the parallel (antiparallel) rotator, where electrons (protons) are collected around the rotation axis and protons (electrons) around the equator plane, a polodial (outward directed) current along the magnetic field lines exists, consisting of electrons (protons). In the equator plane we found  protons (electrons) diffuse out of the simulation volume. In the context of the polodial current in the case of the antiparallel rotator averaged protons energies up to $10^{16}$ eV are found. 

For inclined and orthogonal rotators ($0 < \chi \le \frac{\pi}{2}$) outward directed currents  consisting of electrons along the magnetic field lines are found. The angle range according the polar angle of these currents decreases with decreasing values of $|\chi - \frac{\pi}{2}|$. With inclination angles $30^{\circ} < \chi \le 90^{\circ}$ closed currents consisting of protons are observed in the range of a few $r_N$, starting and ending at the surface of the sphere. Given $\frac{\pi}{2} < \chi \le \pi$ the sign of the particles changes. In the case of the  $120^{\circ}$--rotator, investigated as an example for inclined rotators with $\chi > \frac{\pi}{2}$ and with currents consiting of protons, averaged proton energies up to $10^{17}$ eV has been proven.

The influence of the $[{\bf E, B}]$-drift on the structure of the magnetosphere is increasing  with decreasing $|\chi - \frac{\pi}{2}|$. Particles of both sign (electrons and protons) are streaming due to this force back to the surface of the sphere.   

\section{Discussion}

The resulting quasi-stationary magnetospheres are not global force-free. In regions with vanishing particle densities the vacuum electromagnetic field is approximately undisturbed by the non-neutral plasma. In regions with high particle number densities the  projection of the electric vector onto the tangent to the magnetic field vanishes nearly, but not complete. Small deviations from a force--free situation leads due to the extrem strong electromagnetic fields immediately to high particle energies. All in all, independent of the inclination angle, highly relativistic plasmas are found, which leads to the necessity to take the radiation reaction into account.

In the case of the parallel and antiparallel rotator, due to the fact that we did not found particles with low oder moderate particle energies, the usage of the ultra--relativistic approximation of the Lorentz-Dirac-Landau-equation by \cite{ertl} and \cite{zachariades} in their numerical calcualtion is justifiable, shown by our studies. Futhermore, we can confirm the existence of clouds with different charges, seperated by regions of vanishing particle number density (vacuum--gaps), also found by \cite{krause}, \cite{ertl}, \cite{zachariades}, \cite{wolf94} and \cite{neukirch}. All these works including ours found no global force-free magnetosphere. Force-free magnetospheres are often used as an assumption in analytical works. The structure of the quasi-stationary magnetospheres predicted by \cite{jack1} with analytical models can be confirmed with our studies, especially the importance of the FFS. Nevertheless, the closed polodial currents proposed by \cite{jack1} can not be proven by our studies due to the limited simulation volumen of 20 $r_N$. The density distribution of the plasma is described very well by only one quadrupole mode, confirming with \cite{goldreich} in their amplitude and power law. 

In the case of the inclined and orthogonal rotator only few literature is published. The by \cite{ endean} proposed approach solving the stationary Valsov-equation and the Maxwell-equations selfconsistently led to an inside the light-cylinder corotating, charge seperated magnetosphere. The charge separation independent of the inclination angle was also found by \cite{wolf94}, which can be confirmed by our present studies. Basicly, the structure of the quasi-stationary magnetospheres in our calculations confirms with those found by \cite{jack2} analytically. Furthermore, we can attest the dependency of the electric monopole of the rotating sphere of the declination angle, first proposed by \cite{jack2}. Beside this, for the first time with our work statements on particle engeries, currents and drifts based on numerical, full dynamical studies are possible. Due to the limited simulation volume we can not prove if outside the light-cylinder closed currents are formed (starting and ending on the surface of the sphere). The knowledge of the global current system is important due to the fact that the $[{\bf E, B}]$-drift is important at long time scales. As a consequence of these force charged particles move back to the surface and may change the electric monopol of the sphere. The structure of the FFS and resulting from this the structure of the magenetosphere is depending on the electric monopol of the sphere.

The high particle energies in the existing polodial currents are relevant for induced pair production. In the present paper we were able to ignore the resulting $e^{\pm}$--plasma. Nevertheless, relativistic currents in these plasmas may cause microscopic instabilities which could explain the non--thermal radiation in real pulsar  magnetospheres. Analytical studies on this topic (e.g.~\cite{lyutikov}) assume Lor\-entz--factors in these currents of $\gamma \approx 10^7$ and particle densities in the scale of the  Goldreich \& Julian--density, as were confirmed by our studies.

Regarding the discussion about neutron stars as cosmic accelerators for ultra high energy cosmic ray particles the very high particle energies proven by our studies are remarkably. I.e., in the case of the $120^{\circ}$--rotator we found averaged proton energies up to $10^{17}$ eV. In the case of higher magnetic fields, than given by the 'standard set of parameters' higher particle energies are possible. But, investigating these cases one schould prove whether a classical approach is applicable. It is important to note, that due to the limited simulation volume used in our work we can not predict if these high energy particle are able to leave the neutron star magnetoshphere and if, at which particle energies.

\section{Summary}

In this paper we studied relativistic  magnetospheres of rotating cosmic magnets (neutron stars/pulsars) with arbitrary inclination of the magnetic against the rotation axis. Concentrating on the regime dominated by the force--free surface (FFS) we developed a macroscopic description of a cold, collisionless two-component fluid, consisting of electrons and protons, taken into account the radiation reaction  and carried out selfconsistent numerical calculation of relativistic magnetospheres of neutron stars. 

According the first two moments of the relativistic Vlasov--equation the equation of motion of the fluid components are derived. Under the assumption of a cold, collisionless plasma considering the radiation reaction is possible. Due to missing velocity dispersion of the fluid components the radiation reaction term of the equation of motion for a single, charged particle can be added to the equation of motion of the fluids in the macroscopic description. Dealing with near zone of rotating cosmic objects up 20 sphere radii the influence of existing currents to the vaccum magnetic dipol field can be neglected.  

Beside the investigations regarding the parallel and antiparallel rotator, where we are able to confirm with our present studies many analytical predictions given by other authors in the past, in this paper for the first time the magnetospheres of the more general and complex system of inclinded and orthogonal rotators are investigated in the regime of the force--free surfaces (FFS) by the numerical calculation, using a full dynamical approach and taken into account radiation reaction. As in earlier work of our group, a 'standard set of parameters' is used. 

Under these conditions, the following results are found: Global charge separation exists for all degrees of inclination of the magnetic against the rotation axis with highest particle densities of $10^{12} {\rm cm}^{-3}$. Clouds of different charge are seperated by regions of vanishing particle number density. As expected, test particles inserted into the latter regions propagate into one of the adjacent clouds. The dependency of the electric monopole of the rotating sphere on the inclination angle is given by $q_s = \frac{2}{3}\frac{\mu}{r_L} \cos \chi $. Furthermore strong polodial currents exist and locally averaged particle energies typically range up to $10^{16}-10^{17}$ eV, depending on the inclination angle. 

The results given by the presented work can be used as a starting point for an analytical description of neutron star magnetospheres in the near zone. A suitable approach have to consider a relativistic, non-neutral, charge seperated plasma. Although we prove corotation, an analytical approach should used a splitting in a corotational and non-corotational part of the description. In general, it is not usefull to assume a global force-free magnetosphere. The radiation reaction has to be taken into account, while using the  ultra--relativistic approximation of the Lorentz-Dirac-Landau-equation is appropriate.


\appendix

\section{Integration of the Continuity Equation}
\label{FCT}

This appendix describes the numerical integration of the continuity equation $\partial_{\m}j^{\m} =0$ in detail. The used method is called {\it flux corrected transport} (FCT), and for details regarding general aspects of this method we refer to \cite{zalesek}.

What follow we introduce a 'low-order' and a 'high-order' scheme in three spatial coordinates in order to construct a conservative FCT scheme. A spherical coordinate system is used and $i,j,k$ denote the grid points to the ${\bf e}_r, {\bf e}_{\t}$ and ${\bf e}_{\p} $--direction. 

The flux which flows from the cell $i-1$ to cell $i$ is called $F_{i-\frac{1}{2},j,k}$ and the flux from cell $i$ to cell $i+1$ $F_{i+\frac{1}{2},j,k}$. $G$ and $H$ are the corresponding fluxes in ${\bf e}_{\t}$-- and ${\bf e}_{\p}$--direction. With this notation a conservative discretisation is given by:
\begin{equation}
\r^{n+1}_{i,j,k} = \r^{n}_{i,j,k} 
- F_{i+\frac{1}{2},j,k} + F_{i-\frac{1}{2},j,k}
- G_{i,j+\frac{1}{2},k} + G_{i,j-\frac{1}{2},k}
- H_{i,j,k+\frac{1}{2}} + H_{i,j,k-\frac{1}{2}}\, .
\end{equation}

As the 'low-order' scheme we use the Donor-Cell method in three dimensions
\begin{eqnarray}\nonumber
\label{donor}
\r^{n+1}_{i,j,k} = \r^{n}_{i,j,k}\nonumber  
&-& 0.5(\xi_1-|\xi_1|)\r^n_{i+1,j,k}\,+\,0.5(\xi_1+|\xi_1|)\r^n_{i-1,j,k}\\
&-& 0.5(\xi_2-|\xi_2|)\r^n_{i,j+1,k}\,+\,0.5(\xi_2+|\xi_2|)\r^n_{i,j-1,k}\\ 
\nonumber
&-& 0.5(\xi_3-|\xi_3|)\r^n_{i,j,k+1}\,+\,0.5(\xi_3+|\xi_3|)\r^n_{i,j,k-1}\, ,
\end{eqnarray}
with the Courant-numbers $\xi$ concerning the three spatial coordinates which are indicated by the indices $1,2,3$. The stability condition regarding this 'low-order' scheme is given by: $|\xi_1| + |\xi_2| + |\xi_3| \le 1$. 

A stable 'high-order' scheme (in three spatial coordinates) is developed by \cite{dukowicz}:
\begin{equation}
\label{high-order}
\r^{n+1} = \r^n - \Delta t\, (\Bn,  ( \r {\bf v} - 
0.5 \, \Delta t \,{\bf v} (\Bn ,n {\bf v}) )\,) \,.
\end{equation}

Now we are able to write down the discretisation schemes. Using dimensionless units  $\Delta t^{\prime} = \omega \, \Delta t$ and $\Delta r^{\prime} = r_L^{-1} \, \Delta r$ and supressing the primes. The continuity equation is now given by $ \partial_t N + \partial_i N \b^i = 0 \,$ , where $N$ is the particle density in the inertial frame of reference.
 
The currents regarding the Donor-Cell method are given by
\begin{eqnarray} \nonumber
\fl && (N \b_r)_{i+\frac{1}{2},j,k} = 
 \frac{1}{2}\,[\,(\b_r)_{i+1,j,k} - |(\b_r)_{i+1,j,k}|) N_{i+1,j,k} 
 + \frac{1}{2}\,((\b_r)_{i,j,k} 
+ |(\b_r)_{i,j,k}|\,] N_{i,j,k} \,, \\ \nonumber
\fl && (N \b_{\t})_{i,j+\frac{1}{2},j} =
 \frac{1}{2}\,[\,(\b_{\t})_{i,j+1,k} - |(\b_{\t})_{i,j+1,k}|) N_{i,j+1,k}
+ \frac{1}{2}\,((\b_{\t})_{i,j,k} 
+ |(\b_{\t})_{i,j,k}|\,] N_{i,j,k} \,, \\ \nonumber
\fl && (N \b_{\p})_{i,j,k+\frac{1}{2}} =
 \frac{1}{2}\,[\,(\b_{\p})_{i,j,k+1} - |(\b_{\p})_{i,j,k+1}|) N_{i,j,k+1}
+ \frac{1}{2} \,((\b_{\p})_{i,j,k} 
+ |(\b_{\p})_{i,j,k}|\,] N_{i,j,k}\,.
\end{eqnarray}
Using the following abbreviations for the surface elements $
{\rm Fl}_{r ,\, (i,j,k)} = 2 \, r^2_i \, \sin \t_j 
\sin \frac{\Delta \t_j}{2} \, \Delta \p_k  \, $, 
${\rm Fl}_{\t ,\, (i,j,k)} = \frac{1}{2}\,(r^2_{i+\frac{1}{2}} - 
r^2_{i-\frac{1}{2}} ) \, \sin \t_j 
\Delta \p_k \, $, 
${\rm Fl}_{\p ,\, (i,j,k)} =  \frac{1}{2}\,(r^2_{i+\frac{1}{2}} - 
r^2_{i-\frac{1}{2}} )\, \Delta \t_j $
the fluxes are given by:
\begin{eqnarray}
\label{low-order-flux}\nonumber
 \fl && F_{i+\frac{1}{2},j,k} =
\Delta t \, {\rm Fl}_{r ,\,(i+\frac{1}{2},j,k)}\,\,
[{\rm min}(\,0,(\b_r)_{i+1,j,k}\,)\,N_{i+1,j,k}\, +  {\rm max}(\,0,(\b_r)_{i,j,k}\,)\,N_{i,j,k} ] \,,\\ \nonumber
\fl && G_{i,j+\frac{1}{2},k}  = 
\Delta t \, {\rm Fl}_{\t, \,(i,j+\frac{1}{2},k)} \,
[{\rm min}(\,0,(\b_{\t})_{i,j+1,k}\,)\,N_{i,j+1,k}
+ {\rm max}(\,0,(\b_{\t})_{i,j,k}\,)\,N_{i,j,k} ]\,,\\ \nonumber
\fl && H_{i,j,k+\frac{1}{2}} = 
\Delta t \, {\rm Fl}_{\p ,\,(i,j,k+\frac{1}{2})} 
[{\rm min}(\,0,(\b_{\p})_{i,j,j,k+1}\,)\,N_{i,j,k+1} 
+ {\rm max}(\,0,(\b_{\p})_{i,j,k}\,)\,N_{i,j,k}] \, . \nonumber
\end{eqnarray}
With these fluxes the 'low-order' scheme is given by 
\begin{equation*}
\fl N^{n+1}_{i,j,k} = N^{n}_{i,j,k} 
+ \frac{1}{\Delta V_{i,j,k}}
\left[ - F_{i+\frac{1}{2},j,k} + F_{i-\frac{1}{2},j,k}
 - G_{i,j+\frac{1}{2},k} + G_{i,j-\frac{1}{2},k}
- H_{i,j,k+\frac{1}{2}} + H_{i,j,k-\frac{1}{2}}\right] \, .
\end{equation*}
Referring to (\ref{high-order}) the 'high-order' scheme is given by
\begin{eqnarray}\nonumber
\fl N^{n+1} = N^n - \Delta t \,
&& \left[  \frac{1}{r^2} \partial_r  \left( r^2 
\left( N \b_r - \frac{1}{2} \Delta t\, \b_r \,g
\right) \right) 
+ \frac{1}{r \sin \t} \partial_{\t}\left( \sin \t \,
\left(  N \b_{\t} - \frac{1}{2} \Delta t\, \b_{\t} \, g 
\right)\right) \right.\\ \nonumber
&& \left. + \frac{1}{r \sin \t} \partial_{\p}
\left(  N \b_{\p} - \frac{1}{2} \Delta t\, \b_{\p} \,g \right) \right]\,,
\end{eqnarray}
with $g := \partial_i ( N \b^i) = \frac{1}{r^2}\partial_r ( r^2 N \b_r)
+ \frac{1}{r \sin \t} \partial_{\t} (\sin \t N \b_{\t})
+ \frac{1}{r \sin \t} \partial_{\p} (N \b_{\p})   \,$.
The fluxes regarding the 'high-order' scheme result in:
\begin{eqnarray} \nonumber
\label{f1}
F_{i+\frac{1}{2},j,k} &=& \Delta t {\rm Fl}_{\r,\,(i+\frac{1}{2},j,k)} \,\left[
( N \b_r)_{i+\frac{1}{2},j,k} - \frac{1}{2} \Delta t\,
(\b_r)_{i+\frac{1}{2},j,k} \left[\phantom{\frac{1}{2}}\right. \right. 
\\ \nonumber
&& \frac{1}{r^2_{i+\frac{1}{2}} \, \Delta r}
\left(\,(r^2 N \b_r)_{i+1,j,k}-(r^2 N \b_r)_{i,j,k}\,\right) \\ \nonumber
&+&   \frac{1}{4\,r_{i+\frac{1}{2}}\sin \t_j \Delta \t}
\left[\, \sin \t_{j+1} ( \,(N \b_{\t})_{i,j+1,k} +  (N 
\b_{\t})_{i+1,j+1,k})\right.  \\ \nonumber
&& \hspace{2.8cm} \left. - \sin \t_{j-1} (\, (N \b_{\t})_{i,j-1,k} +
 (N \b_{\t})_{i+1,j-1,k}) \right]\\ \nonumber
&+& \frac{1}{4\,r_{i+\frac{1}{2}}\sin \t \Delta \p}
\left[ \, (N \b_{\p})_{i,j,k+1} +  (N \b_{\p})_{i+1,j,k+1} \right. \\ \nonumber
&&\left. \left. \left. \hspace{2.4cm} 
- \left(\,(N \b_{\p})_{i,j,k-1} + (N \b_{\p})_{i+1,j,k-1}\,\right) \,\,\right] 
\phantom{\frac{1}{2}} \right]\right]\,,
\end{eqnarray}
\begin{eqnarray} \nonumber
\label{f2}
G_{i,j+\frac{1}{2},k} &=&
\Delta t \, {\rm Fl}_{\t,\,(i,j+\frac{1}{2},k)} \!\! \left[
(\sin \t \, N \b_{\t})_{i,j+\frac{1}{2},k} \! - \! \frac{1}{2}
\Delta t\,
(\sin \t \, \b_{\t})_{i,j+\frac{1}{2},k} 
\left[\phantom{\frac{1}{2}}  \right. \right. \\ \nonumber
&& \frac{1}{r \sin \t_{j+\frac{1}{2}}
\Delta \t}
\left[ (\sin \t \, N \b_{\t})_{i,j+1,k}- (\sin \t \, N \b_{\t})_{i,j,k}
\right] \\ \nonumber
&+& \frac{1}{4  \, r^2\,\Delta r}
\left[\, r^2_{i+1}((N \b_{r})_{i+1,j,k} + (N \b_{r})_{i+1,j+1,k})
\right. \\ \nonumber 
&& \left. \hspace{1.3cm} - r^2_{i-1}((N \b_{r})_{i-1,j,k} + (N \b_{r})_{i-1,j+1,k})\, 
\right]\\ \nonumber
&+& \frac{1}{4\,r \sin \t_{j+\frac{1}{2}} \Delta \p}
\left[ \,(N \b_{\p})_{i,j,k+1} +  (N \b_{\p})_{i,j+1,k+1} \right. \\ \nonumber
&&\left. \left. \left.
\hspace{2.6cm} - (\,(N \b_{\p})_{i,j,k-1}+(N \b_{\p})_{i,j+1,k-1}\,)\,\right] 
\phantom{\frac{1}{2}}\right] \right]\,,
\end{eqnarray}
\begin{eqnarray} \nonumber
\label{f3}
H_{i,j,k+\frac{1}{2}} &=&  
\Delta t {\rm Fl}_{\p , \,(i,j,k)} \, \left[
(N \b_{\p})_{i,j,k+\frac{1}{2}} - \frac{1}{2} \Delta t\,
(\b_{\p})_{i,j,k+\frac{1}{2}} \left[ \phantom{\frac{1}{2}}
\right. \right. \\ \nonumber
&&\frac{1}{r \sin \t \Delta \p} \left[ \, 
(N \b_{\p})_{i,j,k+1}-(N \b_{\p})_{i,j,k}\,\right] \\ \nonumber
&+& \frac{1}{4\,r^2\, \Delta r}
\left[\, r^2_{i+1}((N \b_{r})_{i+1,j,k} + (N \b_{r})_{i+1,j,k+1}) 
\right. \\ \nonumber
&& \left.\hspace{1.3cm}  - r^2_{i-1}\left( (N \b_{r})_{i-1,j,k} + (N \b_{r})_{i-1,j,k+1}\, \right)
\right]\\ \nonumber
&+& \frac{1}{4\,r \sin \t \Delta \t}
\left[ \,\sin \t_{j+1}(\,(N \b_{\t})_{i,j+1,k} + (N \b_{\t})_{i,j+1,k+1})
\right. \\ \nonumber
&& \hspace{1.5cm} \left. \left.  \phantom{\frac{1}{2}}
- \sin \t_{j-1}(\, (N \b_{\t})_{i,j-1,k} + (N \b_{\t})_{i,j-1,k+1}\,) ]
\right] \right]\,.
\end{eqnarray}

Using the mentioned fluxes for the 'low-' and the 'high-order' scheme the construction of a full FCT scheme is straightforward (for details we refer to \cite{zalesek}). This FCT scheme is stable if the following conditions are fulfilled: $|\xi_1|^{\frac{2}{3}} + |\xi_2|^{\frac{2}{3}} + |\xi_3|^{\frac{2}{3}} \le 1 \,,\Delta t \le \left[ \left(\frac{|\beta_r|}{\Delta r}\right)^{\frac{2}{3}}+\left( \frac{|\beta_{\t}|}{r \Delta \t}\right)^{\frac{2}{3}}+\left( \frac{|\beta_{\p}|}{r \sin \t \Delta \p}\right)^{\frac{2}{3}}\right]^{-\frac{3}{2}}\,. $


\end{document}